\documentclass[aps,prb,twocolumn,showpacs,floatfix,groupedaddress]{revtex4}
\usepackage{graphicx}
\usepackage{braket}
\usepackage[caption=false]{subfig}
\usepackage{natbib}
\newcommand{\non}{\nonumber}
\newcommand{\bea}{\begin{eqnarray}}
\newcommand{\eea}{\end{eqnarray}}
\newcommand{\beq}{\begin{equation}}
\newcommand{\eeq}{\end{equation}}
\usepackage{color}

\begin{document}

\title{Majorana wavefunction oscillations, fermion parity switches, and disorder in Kitaev chains}

\author{Suraj S Hegde}
\email{shegde2@illinois.edu}
\author{Smitha Vishveshwara}

\affiliation{ Dept.of Physics, University of Illinois, Urbana-Champaign}

\begin{abstract}
We study the decay and oscillations of Majorana fermion wavefunctions and ground state (GS) fermion parity in one-dimensional topological superconducting lattice systems. Using a Majorana transfer matrix method, we find that Majorana wavefunction properties are encoded in the associated Lyapunov exponent, which in turn is the sum of two independent components: a `superconducting component' which characterizes the gap induced decay, and the `normal component', which determines the oscillations and response to chemical potential configurations. The topological phase transition separating phases with and without Majorana end modes is seen to be a cancellation of these two components. We show that Majorana wavefunction oscillations are completely determined by an underlying non-superconducting tight-binding model  and are solely responsible for GS fermion parity switches in finite-sized systems. These observations enable us to analytically chart out wavefunction oscillations, the resultant GS parity configuration as a function of parameter space in uniform wires, and special parity switch points where degenerate zero energy Majorana modes are restored in spite of finite size effects. For  disordered wires, we find that band oscillations are completely washed out leading to a second localization length for the Majorana mode and the remnant oscillations are randomized as per Anderson localization physics in normal systems. Our transfer matrix method further allows us to i) reproduce known results on the scaling of mid-gap Majorana states and demonstrate the origin of its log-normal distribution, ii) identify contrasting behavior of disorder-dependent GS parity switches for the cases of even versus odd number of lattice sites, and iii) chart out the GS parity configuration and associated parity switch points as a function of disorder strength. 
\end{abstract}



\maketitle

\section{Introduction}

Majorana fermions  have become an important and active topic of research over the last decade in condensed matter physics~\citep{Alicea12, Leijnse12, Elliott15}. Defined as their own anti-particles, in condensed matter settings, they have most commonly  been predicted to exist as particular zero energy bound states in fermionic topological superconductors\citep{Kitaev01}. Advances in experimental realizations of such Majorana modes were facilitated by various proposals which involved superconductor-topological insulator interfaces ~\citep{Fu08}, proximity-induced superconductivity in spin-orbit coupled wires~\citep{Oreg10, Lutchyn10}, ferromagnetic atoms in proximity to superconductor ~\citep{Choy11, Nadj-Perge13}. Several reports of experiments show highly suggestive evidence of the existence of the Majorana modes in these systems ~\citep{Mourik12, Deng12, Rokhinson12, Das12, Finck13, Churchill13, Nadj-Perge14}. In addition to the realization of Majorana fermions in and of themselves, conclusive evidence of their existence is of great interest from the perspective of topological quantum computing given their non-Abelian braiding statistics~\citep{Ivanov01} and that they form a natural basis for topological qubits~\citep{Nayak08}.

A key feature of systems hosting zero-energy Majorana modes is the double degeneracy of the ground state due to the pairwise existence of these modes. The Majorana modes at the edges can be combined to form a Dirac fermionic state, which can be either occupied or empty. The two degenerate states are thus characterized by fermion parity. In topological $p$-wave paired superconducting wires, as is our focus here, these Majorana modes typically appear as bound states confined to each end of the wire. For finite-sized wires, the Majorana boundary modes interact via their wavefunction overlap within the bulk. This leads to a splitting in the ground state degeneracy which is exponentially small compared to the superconducting gap ~\cite{Kitaev01}. It has been shown that this splitting in degeneracy is an oscillating function of system parameters,such as chemical potential, superconducting gap and the length of the chain ~\citep{Pientka13, Meng09, Meng10, Prada12, Rainis13, Sarma12, Ben-Shach15, Thakurathi14}. Due to the oscillatory dependence of this splitting, the two low-lying states cross each other at the Fermi energy at regular intervals ~\citep{Meng10, Zyavgin15, Kao14, Hegde14} and thus switch the ground state fermion parity. In a previous collaborative work~\citep{Hegde14}, we mapped out points on the topological phase diagram of a uniform wire where such degeneracy occurs and charted regions characterized by even versus odd fermion parity. These studies show how even in the realistic scenario of wires of finite length, the system can be tuned to restore the double degeneracy at these level crossings. The ability to tune between parity sectors provides a powerful knob in that it effectively corresponds to performing certain non-Abelian rotation operations~\citep{Burello13, Sau11, vanHeck12, Burnell14,Hyart13}.

  Given the importance of accessing Majorana mode related degeneracy points and specific parity sectors in realistic geometries, here we perform a detailed analysis of Majorana wavefunction oscillations and fermion ground state parity properties in finite sized Majorana wires. We model the wire as the oft-used one-dimensional superconducting tight-binding lattice system known as the Kitaev chain. We investigate the features of uniform chain as well as that in the presence of an on-site disorder potential. Focusing on the Majorana end mode in a semi-infinite wire, one naturally finds that its wavefunction is characterized by two features - a decaying envelope and oscillations~\citep{Pientka13, Meng09, Meng10, Prada12, Rainis13, Sarma12, Hegde14}. Borrowing from previous collaborative work by one of us~\cite{DeGottardi13, DeGottardi13A}, we find that these features can be decomposed into two parts - one entirely due to the superconducting pairing potential and another stemming from the underlying normal system in which the pairing potential is absent ~\cite{Motrunich01, Reider13, Sau12}. While both can contribute to the decaying envelope, the oscillations completely reflect those of wavefunctions in the underlying normal system. Thus, in uniform systems, we find that Majorana wavefunctions exhibit oscillations only in a regime where band oscillations of the underlying normal problem are allowed and that this regime is confined to a circular region in the topological phase diagram. This region is intimately connected to the region in which spin-spin correlations show oscillations in the transverse field XY spin chain, a system that directly maps to the Kitaev chain ~\citep{Barouch71}. For the disordered case, band oscillations are washed out. The underlying normal system provides a contribution to the decay as well as random oscillations, both stemming from the behavior of one-dimensional normal state wavefunctions in the presence of a disordered landscape. Thus, in the disordered Majorana wire, Anderson localization physics is crucial in determining the nature of Majorana end mode wavefunction decay and oscillations. 
  
  The behavior of ground state fermion parity  directly depends on the manner in which Majorana end mode wavefunctions oscillate. Consequently, in the uniform case, parity switches only occur within the circle of oscillations; here we map these switches for a finite size wire. We show that this map of the uniform case serves to inform parity switching behavior in the disordered case. Our studies of parity in this disordered case addresses several issues. We establish that zero energy level crossings do indeed correspond to parity switches, a correspondence which is not at all obvious due to a proliferation of low-energy states, unlike in the uniform case. Based on the uniform case analysis, we chart out regions where parity switches are expected to occur, the manner in which they do so, and their dependence on even versus an odd number of lattice sites. These observations are relevant to a slew of parity-based studies and proposals in realistic solid state systems where disorder is a given.

Beyond topological aspects alone, the disordered one-dimensional p-wave superconducting wire has been actively studied for more than a decade under symmetry classification of Class-D due to its localization-delocalization properties characteristics ~\citep{Motrunich01, Brouwer00, Brouwer03, Gruzberg05}. One of the highlighting features of this system is the existence of a delocalised multi-fractal wavefunction  at a critical point and the surrounding `Griffiths phase'. In this phase a proliferation of the low energy bulk states into the superconducting gap causes the density-of-states to diverge at zero energy. This critical point in fact separates the topological and trivial phases; the delocalized state at the critical point provides a channel for the Majorana end mode in the topological phase to vanish in the trivial phase. Subsequent to pioneering work in Ref.\onlinecite{Motrunich01}, several recent works have studied Majorana physics in the presence of disorder ~\cite{ Reider14, Reider13, Reider12, Pientka12, Brouwer11, Neven13, Bagrets12, Hui14, Lobos12, Sau13A, Fregoso13, Stanescu12, Beenakker14, Pikulin12, Fulga11, Akhmerov11, Beenakker13, DeGottardi13, DeGottardi13A}.  Along with some other random-matrix based treatments~\citep{Beenakker13, Beenakker13A}, our present work is one of the few to explicitly address the issue of Majorana mode based fermion parity switches in the presence of disorder. 

Finally, comments on our formalism are in order, particularly with regards to the technique of transfer matrices.  Transfer matrices  have been used extensively in the context of disorder as a means of studying the manner in which wavefunctions behave across a system and of extracting properties such as transmission co-efficients, localization lengths, and conductance. It has also been employed in the context of topological phases ~\cite{Hatsugai93, Dwivedi15} In class D systems, transfer matrix method has proved highly useful in deriving their unusual behavior and critical features ~\cite{Gruzberg05, Ludwig12, Ryu12, Brouwer05}. With regards to Majorana mode physics, transfer matrices have been employed to define a topological invariant based on normalization properties of the Majorana wavefunction and identify the topological phase diagram in the presence of various potential landscapes including that of disorder ~\citep{DeGottardi13A, Mandal15, Fregoso13, Sau12}. In our analyses, the method provides a direct way of treating the above-mentioned separation of the Majorana wavefunction in terms of the superconducting and normal contributions as well as identifying the presence of zero energy states in finite length wires. In the presence of disorder, the technique provides us a transparent way of linking Anderson localization physics and Majorana wavefunction properties.

 The presentation of the paper is as follows: In Section \ref{Kitchain}, we review the salient features of the Kitaev chain Hamiltonian, its topological phases, and the formalism of transfer matrices for describing Majorana zero modes. In Section \ref{WaveOsc}, using the transfer matrix formalism, we outline generic features of Majorana wavefunctions in semi-infinite wires and give detailed descriptions for the uniform and disordered cases. In Section ~\ref{ParSwitch}, we focus on finite-sized systems and obtain a general condition for the occurrence of parity switches in terms of Majorana transfer matrices. We also map out the fermion ground state parity in the topological phase diagram for the uniform case. In Section ~\ref{DisParity} we study the effect of disorder on degeneracy-splitting and parity switches.

\section{Kitaev chain and transfer matrix set-up}
\label{Kitchain}
In this section, we begin by introducing a lattice version of the most basic Majorana wire, namely the Kitaev chain Hamiltonian, as the starting point of our studies. The Hamiltonian allows for the existence of i) a topological phase that hosts zero-energy Majorana fermionic bound states at the ends of the chain giving rise to a doubly degenerate ground state and ii) a non-topological phase where these states are absent. We then review a transfer matrix formalism and the associated Lyapunov exponent description for extracting the localization behavior of Majorana mode eigenstates. Finally we also introduce a map that separates contributions of the superconducting order from the underlying normal state properties to the Lyapunov exponent. This set-up forms the basis of our studies in subsequent sections.

{\it Kitaev chain Hamiltonian. ---}

 The prototypical model for studying topological $p$-wave superconductivity in one-dimension is the Kitaev chain \citep{Kitaev01}. This lattice model consists of non-interacting spinless fermions on each site having nearest neighbor tunneling of strength $w$, nearest neighbor superconducting pairing of strength  $\Delta$, and an on-site chemical potential $\mu_n$.Its associated Hamiltonian takes the form
\begin{equation}
H=  \sum_{n=1}^{N-1} (-w c^{\dagger}_{n+1}c_n + \Delta c^{\dagger}_{n+1}c^{\dagger}_n + h.c)- \sum_{n=1}^{N} \mu_n(c^{\dagger}_n c_n-1/2),
\end{equation}
where $h.c.$ denotes Hermitian conjugate.
Here, the $c^{\dagger}_{n}$ and $c_n$ operators represent the creation and annhilation of electrons on site $n$, respectively. Alternatively, these operators can be expressed in terms of a pair of Majorana fermion operators $\hat{a}_n$ and $\hat{b}_n$, namely, $\hat{a}_n=c_n+c_n^{\dagger}$ and $\hat{b}_n=i(c_n^{\dagger}-c_n)$. The Majorana operators satisfy the relations $\hat{a}^{\dagger}_n=\hat{a}_n$, $\hat{b}^{\dagger}_n=\hat{b}_n$ and $\{\hat{a}_n,\hat{a}_m\}=\{\hat{b}_n,\hat{b}_m\}=2 \delta_{mn}$. In terms of the Majorana operators, the Hamiltonian is given by

\bea \non
H_M&=&-\frac{i}{2} \sum\limits_{n=1}^{N-1}\bigg[(w-\Delta) \hat{a}_n \hat{b}_{n+1}
- (w+\Delta) \hat{b}_{n} \hat{a}_{n+1} \bigg] \\
&& - \sum\limits_{n=1}^{N} \frac{i\mu_n }{2} \hat{a}_{n} \hat{b}_{n}. \label{majoranaH} \eea

In the case of a homogeneous chain, $\mu_n=\mu$ for all $n$. The system is in the topological phase for the parameter range $-2w<\mu<2w$ for non-zero superconducting gap. The gap closes at the transition to the trivial phase at $|\mu|=2w$. In the presence of a potential energy landscape, $\mu_n$ exhibits spatial variation and can drastically alter the qualitative features of the phase diagram. Specific to later sections, in the presence of disordered landscapes, the system can even become gapless for a broad region around the transition between the topological and non-topological phase. Previous studies have explored the Kitaev chain and the associated topological phase diagram in the presence of a variety of potential landscapes including (quasi)periodic potentials and disorder \cite{DeGottardi13}.

 {\it Majorana Transfer Matrix and Lyapunov Exponent. ---}
 \label{sec:Transfer}
  The transfer matrix method is designed to study the manner in which wavefunctions propagate through the length of a system. It thus offers a natural means of probing localisation aspects of Majorana end modes and the topological characteristics of the Kitaev chain. In previous works \onlinecite{ DeGottardi11, DeGottardi13}, one of the present authors and collaborators developed Majorana transfer matrix formalism  for determining the topological invariants and charting the phase diagram of the Kitaev chain in the presence of different potential landscapes. 
  
  Here we briefly recapitulate this Majorana transfer matrix technique and the associated Lyapunov exponent description. Given the Majorana Hamiltonian of Eq.\ref{majoranaH}, we first obtain the zero energy Heisenberg equation of motion $[\hat{a}_n,H_M]=0$ for the Majorana operators $\hat{a}_n$. Using this, we obtain the equation for Majorana wavefunctions $a_n$ as :
 \begin{equation}
(w+\Delta) a_{n+1} + (w - \Delta)a_{n-1} + \mu a_n =0.
\label{weqa}
\end{equation}
(Note that the `$a$' without a hat is the Majorana wavefunction and not the operator itself.)
For $2 \leq n \leq N-1$, the modes at different sites are thus related  by the transfer matrix:
\bea \left( \begin{array}{c}
a_{n+1} \\
a_n \end{array} \right) &=&~ A_n \left( \begin{array}{c}
a_n \\
a_{n-1} \end{array} \right), \non \\
\mbox{where}~~ A_n &=& \left( \begin{array}{cc}
-\frac{\mu_n}{w + \Delta} & - \frac{w - \Delta}{w + \Delta} \\
1 & 0 \end{array} \right). \label{tmatrix} \eea
  The transfer matrix for $b_n$ modes have an identical structure except with a change in the sign in $\Delta$. From this point on, unless specified we set $w=1$. We only consider $\Delta$ positive. Our results for the negative half of the phase diagram are the same except that the roles of $a$ and $b$ are switched. The full chain transfer matrix is given by an ordered product of all the transfer matrices from the first site to the last: $\mathcal{A}= \prod_n A_n$. The eigenvalues of the full transfer matrix, $\lambda_{\pm}$,  determine features of the  Majorana mode wavefunctions, such as oscillation and decay.   The eigenvalues can be employed to construct a topological invariant which determines whether or not the system is in the topological phase based on Majorana wavefunction normalizability \citep{DeGottardi11}. Specifically, if the number of eigenvalues within the unit circle of a complex plane is even, then the system is in a topologically non-trivial phase.  
  
  A useful quantity to extract from transfer matrices is the Lyapunov exponent (LE), which in most cases relates to the inverse localization length of the corresponding wavefunction. For the purpose of analyzing a Majorana bound state at one end of the Kitaev chain, we study the Lyapunov exponent for a semi-infinite system having a boundary at one end, defined as 
\beq
\gamma(\mu,\Delta)= \lim_{N \rightarrow \infty}\frac{1}{N} \text{ln}[|\lambda|]
\label{LE}
\eeq
The largest of the two eigenvalues $\lambda_{\pm}$ is taken for the definition. 
It was shown in Ref.~\onlinecite{DeGottardi11} using wavefunction normalization properties that the LE is able to detect a topological phase transition in the Kitaev chain and was used to chart the phase diagram for various potential landscapes i.e with different configurations of $\mu_n$. It was shown that in the topological phase the LE is always negative and a topological phase transition occurs when it crosses zero to become positive. Thus the loci of points of zero LE give the phase boundary.  Here, we will employ the LE not only as a probe of this topological phase transition but also to investigate detailed features of the Majorana wavefunction.
 
 In obtaining the LE for a generic potential landscape, it was recently found that the Majorana transfer matrix could be mapped to that of the transfer matrix of a normal system without a superconducting gap. The map proved powerful in that knowledge of normal state wavefunction properties immediately led to those of the topological superconducting chain. This map is achieved through the following similarity transformation on the transfer matrix in Eq.[~\ref{tmatrix}]~\citep{DeGottardi13}, defined for $0 < \Delta < 1$:
  \beq
  A_n = \sqrt{l_{\Delta}}S \tilde{A_n}S^{-1}
  \eeq
  where $S= diag( l_{\Delta}^{1/4},1/l_{\Delta}^{1/4})$ and $l_{\Delta} = \frac{1-\Delta}{1+\Delta}$. The matrix $\tilde{A_n}$ is  the transfer matrix for a normal tight-binding model in the absence of a  superconducting gap (Note that $w=1$ compared to Eq.\ref{tmatrix}). Its on-site chemical potential terms are rescaled by the transformation $\mu_n \rightarrow \mu_n / \sqrt{1-\Delta^2}$ . Explicitly, the full chain transfer matrix is given by 
  \beq
  \mathcal{A}(\mu_n,\Delta)= \bigg(\frac{1-\Delta}{1+\Delta} \bigg)^{N/2} S \tilde{\mathcal{A}}(\mu_n/\sqrt{1-\Delta^2},\Delta=0)S^{-1}
  \label{simtrans}
\eeq  
This map allows the Lyapunov exponent to be written as a sum of two components, $\gamma(\mu, \Delta)= \gamma_S + \gamma_N$, one that depends purely on the superconducting gap $\gamma_S(\Delta)$ and the other corresponding to the underlying normal tight-binding model $\gamma_N (\mu/\sqrt{1-\Delta^2})$.  We note that such a splitting of Lyapunov exponent was already hinted in Ref. \onlinecite{Motrunich01} and the mapping to normal system in the context of scattering matrices in Ref.\onlinecite{Reider13}

In the following sections, we will employ the Majorana transfer matrix, the Lyapunov exponent and the mapping to a normal system to study the oscillations in Majorana wavefunctions and associated fermion parity switches in both uniform and disordered chains.

\section{Majorana wavefunctions and Oscillations}
\label{WaveOsc}
In this section, we first provide a generic description of the Majorana end modes in terms of their decay and oscillation. We pinpoint the separate contributions due to superconductivity and the presence of a potential landscape for each of these features in terms of the Lyapunov exponent. We use this knowledge to analyze Majorana wavefunctions in the uniform and disordered cases.

\subsection{Generic Formulation}
\label{sec:gen}
The generic transfer matrix given in Eq.~\ref{tmatrix} directly provides information on the Majorana end mode decay profile and oscillatory behavior. Given this matrix for a single slice of the chain, the full-chain transfer matrix is given by $\mathcal{A}= \prod_n A_n$. The eigenvalue equation for the transfer matrix is given by
\beq
\lambda^2+ \text{Tr}(\mathcal{A})\lambda + \text{det}(\mathcal{A})=0
\eeq
Given that $\text{det}(A_n)= (\frac{1-\Delta}{1+\Delta})$, we have $\text{det}(\mathcal{A})=\prod_{n}^{N} \text{det}(A_n)= (\frac{1-\Delta}{1+\Delta})^N$. Thus, the two eigenvalues of the full transfer matrix take the form
\beq
\lambda_{\pm}= \frac{\text{Tr}\mathcal{A}}{2} \pm \sqrt{\bigg(\frac{\text{Tr} \mathcal{A}}{2}\bigg)^2-\bigg(\frac{1-\Delta}{1+\Delta}\bigg)^N}= e^{\pm i\beta}\bigg(\frac{1-\Delta}{1+\Delta}\bigg)^{N/2}.
\label{eigval}
\eeq
Here, the phase $\beta$ can only be real or imaginary depending on the value of $\text{Tr}\mathcal{A}$ given its structure, 
  \beq
  \beta = \tan^{-1}\bigg(\frac{\sqrt{(\frac{1-\Delta}{1+\Delta})^N-(\frac{\text{Tr}\mathcal{A}}{2})^2}}{\frac{\text{Tr}(\mathcal{A})}{2}}\bigg).
\label{beta} 
  \eeq
This phase $\beta$, which is a function of $\mu$,$\Delta$ and $N$ plays an important role in determining the nature of the Majorana wavefunction.

One can see that if $\beta$ is real, the eigenvalues are complex and the wavefunctions have an oscillatory component in addition to the exponentially decaying envelope $(\frac{1-\Delta}{1+\Delta})^{N/2}$. When $\beta$ becomes purely imaginary, then the eigenvalues are purely real, corresponding to overdamped wavefunctions. The specific conditions for the phase $\beta$ to change from real to imaginary depends on the specific potential landscape. In general, the phase $\beta$ alone tracks the response to different potential landscapes, while the scaling factor $(\frac{1-\Delta}{1+\Delta})^{N/2}$ is responsible for the localization of the Majorana mode irrespective of the potential $\mu_n$.

Now, using the similarity transformation Eq.\ref{simtrans}, we can see that this phase factor is completely determined by a simple underlying tight-binding problem that lacks a superconducting gap.  The transformed transfer matrix for a single $\mu_n$ can be easily seen to be a simple tight-binding model as follows:
 \bea 
 \tilde{A}_n &=& \left( \begin{array}{cc}
-\frac{\mu_n}{\sqrt{1-\Delta^2}} & - 1 \\
1 & 0 \end{array} \right).
\label{transnormal} \eea
The Heisenberg equation for the wavefunction $\tilde{a}_n$ corresponding to the above transfer matrix equation is explicitly of the normal tight-binding form:
\beq
(\tilde{a}_{n+1}+\tilde{a}_{n-1})+ \frac{\mu_n}{\sqrt{1-\Delta^2}}\tilde{a}_n=0.
\label{anderson}
\eeq

 Since the similarity transformation $S$ is purely a function of $\Delta$ and does not depend on the individual $\mu_n$, the full chain transfer matrices respect the relationship
\beq
\prod_n A_n = (\sqrt{l_{\Delta}})^N S (\prod_n \tilde{A_n}(\mu_n/\sqrt{1-\Delta^2})) S^{-1}
\eeq
 Since the similarity transformation preserves the trace, the full chain transfer matrix traces are related as
\beq
 \text{Tr}(\mathcal{A})= (\sqrt{l_{\Delta}})^N \text{Tr}(\tilde{\mathcal{A}})
 \eeq
where $\tilde{\mathcal{A}}$ is the full chain transfer matrix for the underlying normal tight-binding model.
 Using this identity in the expression for $\beta$ in Eq.\ref{beta}, we obtain
\beq
\beta= \tan^{-1}\bigg(\frac{\sqrt{1-(\frac{\text{Tr}\tilde{\mathcal{A}}}{2})^2}}{\frac{\text{Tr}(\tilde{\mathcal{A}})}{2}}\bigg)
\label{normalbeta}
\eeq

{\it Thus we make the observation here that the phase factor $\beta$ of the Majorana transfer matrix is solely determined by an underlying normal tight-binding model without a superconducting gap, but with a scaled on-site chemical potential $\mu_n/\sqrt{1-\Delta^2}$. Therefore, oscillations of the Majorana wavefunction and the response to the specific potential landscape are completely determined by properties of the underlying normal state chain. }

 It is precisely these oscillations in Majorana zero modes which determine the degeneracy-splitting  of the ground state in a finite sized Kitaev chain and associated fermion parity switches. Further they also have direct bearing on the oscillations in the spin-spin correlation functions of transverse field XY spin chain, to which there is an exact mapping from the Kitaev chain. The fact that these oscillations can be obtained from a simple tight-binding model easily enables one to extend the study to disorder, periodic and quasi-periodic potential landscapes.

 Further, the division of Majorana wavefunction into the two features of oscillations and decay manifests as an additivity in the Lyapunov exponent (LE). While this splitting of the Lyapunov exponent was pointed out in previous works ~\citep{DeGottardi13, Motrunich01}, here we show the relevance of such a splitting in studying the nature of the Majorana wavefunction.  Specifically, we invoke the definition of the LE in Eq. \ref{LE}, $\gamma(\mu,\Delta)= \lim_{N \rightarrow \infty} \text{ln}[|\lambda|]/N$. Now referring back to the expression for the eigenvalues in Eq.~\ref{eigval}, we consider the two cases of purely real and purely imaginary $\beta$:

(i) When $\beta$ is real, $|\lambda_{\pm}|= (\frac{1-\Delta}{1+\Delta})^{N/2}$. Therefore, the LE is only given by $\gamma_S= -\frac{1}{2}\text{ln}(\frac{1+\Delta}{1-\Delta})$. We call $\gamma_S$ as the `superconducting' component of the LE.  In this case the LE is always negative assuming a finite, positive superconducting gap. This implies that the system is in the topological phase. Moreover, $\beta$ gives rise to an oscillatory piece in the Majorana wavefunction.

For a small superconducting gap, on expanding in terms of small $\Delta$, it can be seen that $\gamma_S \sim -\Delta$. In the continuum limit, the wavefunction of the Majorana mode has an exponentially decaying envelope $e^{-x/\xi}$, where $\xi \sim 1/\Delta$ is the superconducting coherence length which characterizes the localization of the Majorana mode deep in the topological phase. Thus the superconducting component of the LE corresponds to the localization of the Majorana mode at the edge, protected by the superconducting gap. These localization features are immune to any  perturbations that do not close the gap.

 (ii) When $\beta$ is purely imaginary, the phase factor $e^{i \beta}$ is real. As a result, the LE contains two terms. Since  $|\lambda_{\pm}|= e^{\mp |\beta|}(\frac{1-\Delta}{1+\Delta})^{N/2}$, the LE is given by (considering only the largest eigenvalue): 
\bea \non
\gamma &=& \lim_{N \rightarrow \infty}\frac{1}{N} \text{ln}|\text{exp}(i\beta(\mu,\Delta,N))| -\frac{1}{2}\text{ln}(\frac{1+\Delta}{1-\Delta})  \\
 &=& \lim_{N \rightarrow \infty}\frac{1}{N} \beta(\mu,\Delta,N) -\frac{1}{2}\text{ln}(\frac{1+\Delta}{1-\Delta})  \\
 &=& \gamma_N + \gamma_S
 \label{LE2}
\eea

The LE is thus a sum of two components $\gamma_N + \gamma_S$, where $\gamma_N$ is the `normal component' and $\gamma_S$ is the `superconducting component' discussed in case i) above and in Ref.\onlinecite{DeGottardi13}. The `normal component' takes the form $\gamma_N = \lim_{N \rightarrow \infty}\frac{1}{N} \text{ln}|\text{exp}(i\beta(\mu,\Delta,N))|$; it corresponds to the LE of a one-dimensional normal tight-binding model having a general potential landscape represented by a scaled on-site chemical potential ~\cite{Motrunich01}.  To be specific, the tight-binding problem we are considering now contains onsite terms $\mu_n$, which are scaled by a factor involving $\Delta$ , thus giving a LE of the form $\gamma_N(\mu_n/\sqrt{1-\Delta^2},0,N)$ \cite{DeGottardi13}. For a given $\Delta$ and number of lattices sites,  $N$, in the Kitaev chain, we thus need only solve the underlying normal tight-binding problem. While the `superconducting' component is always present in the LE in the presence of a gap, the `normal component' depends on the specific potential landscape under consideration. In the context of disorder, this enables one to use known results from the vast literature of Anderson localization to readily comment on the features of Majorana modes in Kitaev chain.

\subsection{Homogeneous wire: Circle of Oscillations}
\begin{figure}[]
\centering 
\subfloat[][]{\includegraphics[width=0.4\textwidth]{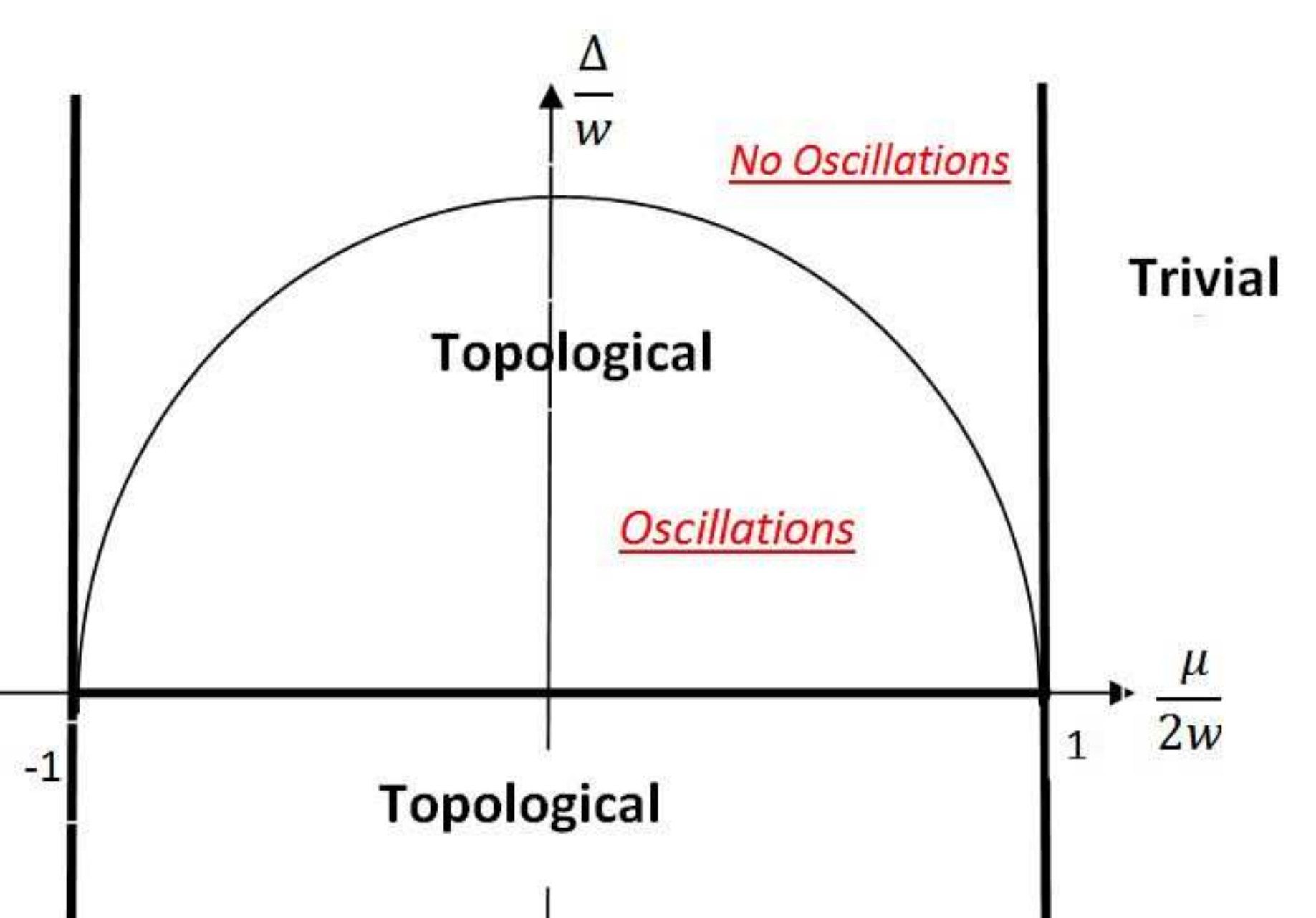}} 
\caption { The topological phase diagram for the uniform Kitaev chain as a function of superconducting gap $\Delta/w$ and chemical potential $\mu/2w$. The focus here is the circle of oscillations (COO) [$\mu^2/4w^2+\Delta^2/w^2 = 1$] within each topological phase marking the boundary across which the nature of Majorana wavefunctions changes. Within the circle, the wavefunction has oscillations under the decaying envelope whereas they are absent outside the circle.}
 \label{fig:betacircle}
\end{figure}

In the case of a homogeneous wire in which the the on-site chemical potential takes on the same value on each site, we can exactly analyze features of the previous sub-section concerning Majorana wavefunction decay and oscillations. The exact wavefunction can be obtained by solving the difference equation Eq.\ref{weqa} by using the Z-transform method, which is an equivalent of Laplace transform for functions of discrete variables. 
For the homogeneous case, the equation of motion is a second order difference equation having constant co-efficients :
\beq
(1+\Delta) a_{n+1} + (1 - \Delta)a_{n-1} + \mu a_n =0.
\eeq
The solution to the above equation proceeds by introducing a power series
\beq
 A(z) = \sum_{n=0}^{\infty} z^{-n} a_n \equiv \mathcal{Z}[a_n],
 \eeq
 where $z$ is a complex variable. The function $A(z)=\mathcal{Z}[a_n]$ is called the Z-transform of $a_n$. Taking the Z-transform of the above difference equation and using properties such as : $\mathcal{Z}\{a_{n-1}\}=z^{-1}A(z)$, $\mathcal{Z}\{a_{n+1}\}= zA(z)-za_0$,($a_0$ is a constant determined by boundary conditions) one can obtain a closed form expression for the Z-transform $A(z)$, given by:
\beq
A(z)= \frac{a_0 z^2}{z^2+ \frac{\mu }{1-\Delta}z+\frac{1+\Delta}{1-\Delta}}
\eeq 
 This Z-transform has a unique inverse, which is the exact solution to the difference equation. Thus the obtained wavefunction is of the form
\begin{equation}
a_n = a_0 C^{n} \bigg[\cos(\beta n)+ \frac{1}{\tan\beta}\sin(\beta n) \bigg]. 
\label{alphan}
\end{equation}
Here, the constant $C=(\frac{1-\Delta}{1+\Delta})^{1/2}$ explicitly reflects the superconducting component and $\beta= \arctan{\frac{\sqrt{4-4\Delta^2-\mu^2}}{\mu}}$ the phase in Eq.\ref{beta}. Thus, as shown in the phase diagram of Fig.  ~\ref{fig:betacircle}, the phase $\beta$ takes on real values only within the circular regime $\mu^2 < 4(1-\Delta^2)$. We call this regime the circle of oscillations (COO); the reality condition on the phase $\beta$ renders the Majorana wavefunctions oscillatory. Outside this regime, the oscillating terms of Eq.\ref{alphan} become hyperbolic and the modes, instead of oscillating, become overdamped\cite{Hegde14, Niu12}. 

This circle of oscillations (COO) has also long been identified in the context of the transverse-field XY spin chain, to which the Kitaev chain can be exactly mapped using the Jordan-Wigner transformation. This circle, termed as the disorder circle, separates the regime of oscillations in spin-spin correlation functions from the regime of no oscillations\cite{Barouch}. Exactly on the circular locus, the ground state becomes separable and can be expressed as a direct product state \cite{Kurmann82, Giampolo09}. As a result, certain entanglement measures, such as the global geometric entanglement, vanish on this locus \cite{Wei11}. Another measure, the `entanglement range', diverges and reflects a change in the pattern of entanglement across this circle \cite{Amico06}. There are similar oscillations in the entanglement spectrum within the circle \cite{Giampolo13}. These aspects in spin chains might have some bearing on the features of Majorana modes. In fact some aspects of Majorana physics can be used to obtain results in spin chains very easily, which otherwise involve cumbersome calculations, as already pointed out in Ref.\onlinecite{DeGottardi11}.

An understanding of the nature of the wavefunctions oscillations can be obtained using the underlying tight-binding problem resulting from the similarity transformation of Eq.\ref{simtrans}. For the homogeneous case, the equation of motion for the underlying normal model takes the form 
\beq
w(\tilde{a}_{n+1}+\tilde{a}_{n-1})+ \frac{\mu}{\sqrt{1-\Delta^2}}\tilde{a}_n=0
\label{normaloscillations}
\eeq
Here $\tilde{a}_n$ is the wavefunction describing the Majorana mode under the envelope coming form the gap $\Delta$ and we have re-inserted the hopping amplitude $w$. Solutions to this equation have the plane-wave form $\tilde{a}_n=D_n e^{\pm i kn}$. Simplifying the equation, we get  $2w\cos k+\frac{\mu}{\sqrt{1-\Delta^2}}=0$ . This relationship imposes the condition that the modes oscillate/propagate only in the region
\beq
 \frac{\mu}{\sqrt{1-\Delta^2}} < 2w
 \eeq
If the above condition is not satisfied, the solutions lie outside the band of propagating modes and are purely decaying.
 Now recasting the above condition, we obtain the relationship : $\mu^2/4 +\Delta^2=1$, which is precisely the equation for the circle of oscillations(COO). Thus, mapping the original Majorana problem to an underlying tight-binding problem, we obtain a picture of the mechanism responsible for oscillations in Majorana modes and how these oscillations immediately vanish outside the circle. 

\begin{figure}[]

\includegraphics[width=0.5\textwidth]{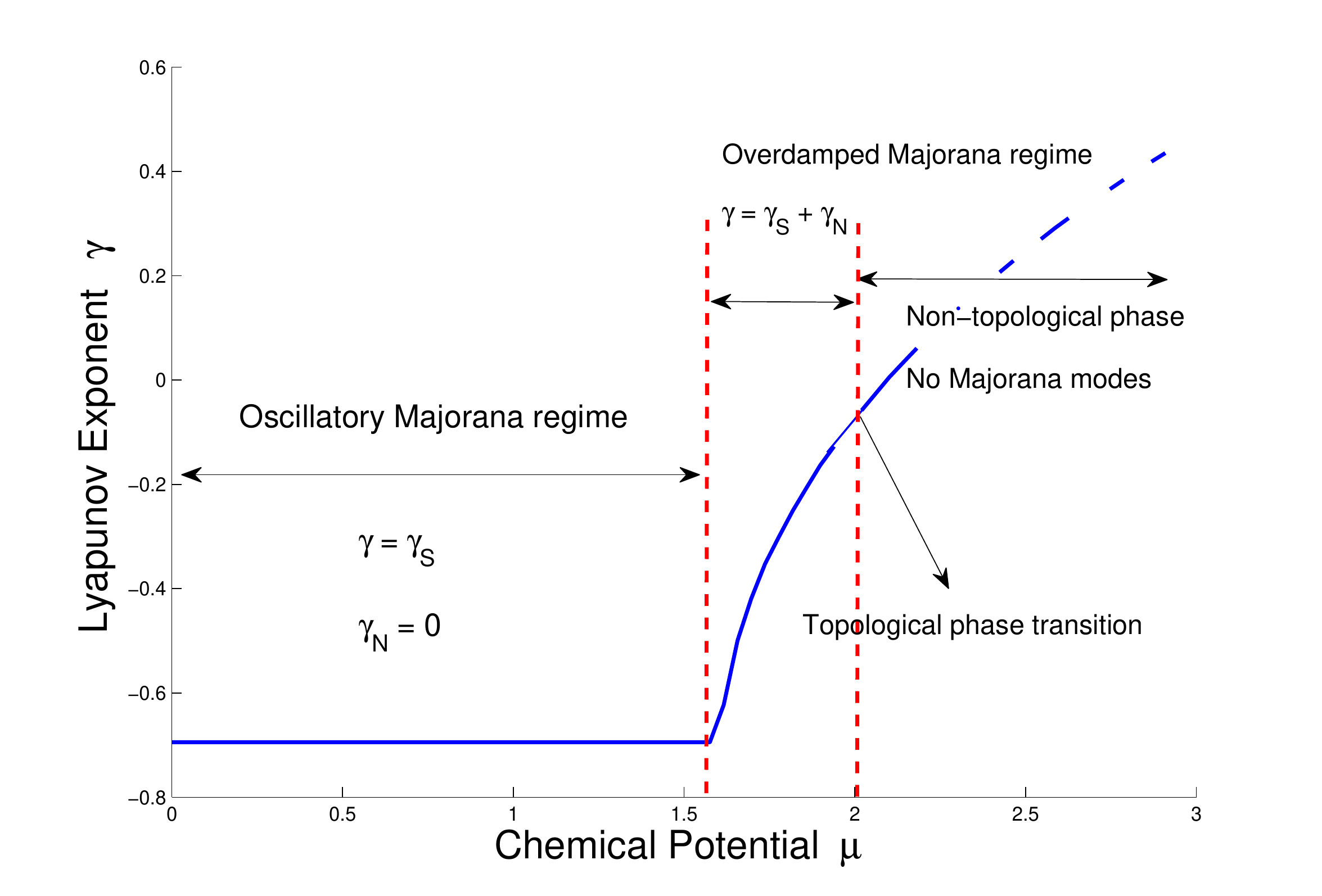}

\caption {The homoeneous Kitaev chain Lyapunov exponent(LE), $\gamma$, measuring Majorana end mode decay as a function of chemical potential for fixed superconducting gap ($\Delta=0.6$). The LE is a sum of a normal and a superconducting component $\gamma = \gamma_N+\gamma_S$, and is a constant within the circle of oscillations since there  $\gamma_N=0$ and $\gamma_S$ is constant for fixed $\Delta$. On crossing the circle, $\gamma_N$ becomes a non-zero increasing function of $\mu$ and ultimately cancels $\gamma_S$, resulting in a zero LE, at the topological phase transition at $\mu=2$. }
 \label{fig:lypo}
\end{figure}

This change in the nature of Majorana oscillations is also tracked by the two components of the Lyapunov exponent.  As discussed in the previous subsection,  when $\beta$ is real i.e everywhere within COO , $\gamma_N = \text{ln}(1)=0$ and the localization of the Majorana mode is only due to the superconducting component $\gamma=\gamma_S =\frac{1}{2}\text{ln}(\frac{1-\Delta}{1+\Delta})$. If $\Delta$ is kept constant, then the LE and thus the localization length are also constant, independent of the chemical potential $\mu$ as shown in Fig.\ref{fig:lypo}. Hence, in this region, the Majorana wavefunction shows an oscillation having an associated wave vector $k_\beta$ and a decay length $\xi_S$ given by
\bea
k_{\beta} & = & \beta  =  \arctan{\frac{\sqrt{4-4\Delta^2-\mu^2}}{\mu}}, \\ \non
\xi_S & = & 1/\gamma_S  =  2/\text{ln}(\frac{1-\Delta}{1+\Delta}),
\eea
respectively. 
 Outside the circle, as discussed earlier, the oscillations disappear and $\beta$ becomes imaginary. Consequently, $\gamma_N$ becomes a non-zero positive number and provides a second localization length. The Majorana wavefunction thus decays over a length scale given by
 \beq
(\gamma_S+\gamma_N)^{-1} = \xi .
\eeq
 
 This expression remains valid within the topological phase; upon encountering the phase boundary between the topological and non-topological phase, the decay length diverges.
 
{\it Topological phase transition. ---} 
The fate of the Majorana wavefunction upon encountering the phase boundary can be studied, for instance, by fixing $\Delta$ and increasing $\mu$, as shown in the diagram of Fig.\ref{fig:lypo}. If we  start deep within topological phase, $\gamma_N$ starts off as zero within the circle, becomes non-zero outside and increases until it becomes equal to $\gamma_S$. This happens exactly at the topological phase transition $\mu=2$, thus resulting in a vanishing of the total LE, indicating complete delocalization of the Majorana mode.

  Since $\gamma_S$ is always non-zero, such delocalization in a generic potential landscape is possible only through its cancellation with $\gamma_N$. Thus the condition for a topological phase transition in general {\it for any potential landscape} obeys:  $|\gamma_S|=|\gamma_N|$.

\subsection{Wavefunctions in the disordered Kitaev chain}
\label{sec:wavefun_disorder}
We now consider the situation in which the on-site chemical potential, $\mu_n$, in the Kitaev chain described by the Hamiltonian in Eq. \ref{majoranaH} exhibits spatial variations. As with normal Anderson localized systems, these variations would reflect a disordered potential landscape. Therefore, the values $\mu_n$ satisfy a typical random distribution, for instance, a Lorentzian, Gaussian or box distribution of an energy scale width $W$. While our discussion on the disordered Kitaev chain is a natural sequel to the above sections, the system itself forms a major topic of study; we save a brief description of this background to a later section.

{\it Additional localization due to disorder.---}
 The addition of Lyapunov exponents in Eq.\ref{LE2} allows us to immediately deduce the effect of disorder in $\mu_n$ on the nature of the Majorana wavefunction. The disordered Majorana wavefunction decays over a length scale given by
\beq
\xi_{dis} = (\gamma_S + \gamma_N)^{-1}.
\label{xi_dis}
\eeq
 As we saw in Subsection.\ref{sec:gen}, $\gamma_S$ is not affected directly due to the variation of $\mu$ and is equal to $-\frac{1}{2}\text{ln}(\frac{1+\Delta}{1-\Delta})$. As long as there is a finite superconducting gap, there is always a corresponding localization scale for the Majorana mode. As for the contribution from $\gamma_N$, this stems from the underlying normal tight-binding model having a scaled on-site disorder potential $\mu_n/\sqrt{1-\Delta^2}$, in other words, a scaled Anderson model in one-dimension. In contrast to the uniform chain, the phase factor $\beta$ in Eq. \ref{eigval} is always imaginary and thus there never exists a region in the phase diagram where $\gamma_N=0$. This observation is consistent with the absence of translational invariance and associated band oscillations and hinges on the well known statement that all states in the Anderson model are localized in dimensions less than two~\cite{Abrahams79}. The behavior of $\gamma_N$ is thus dictated by the localization scale of the disordered wavefunction at zero energy (with respect to the Fermi energy) in the Anderson localization problem and can be studied by invoking the exhaustive literature on the Anderson problem.

As a simple example, consider the case of `Lorentzian disorder'of strength $W$ in which the values of $\mu_n$ are taken from a probability distribution of the form $P(\mu; W) = \frac{1}{\pi}\frac{W}{\mu^2+W^2}$ . The value of the Lyapunov exponent for such a distribution, assuming the scaled chemical potential configuration appropriate for the Kitaev chain,  is given by the form originally derived by Thouless ~\citep{Thouless72}:
 \beq
 \gamma_N \bigg(\frac{W}{\sqrt{1-\Delta^2}},0\bigg)= \text{ln}\bigg(\frac{W}{2\sqrt{1-\Delta^2}} + \sqrt{1+\frac{W^2}{4(1-\Delta^2)}}\bigg)
 \label{LELorentz}
 \eeq
Thus, the normal component of the Lyapunov exponent is a non-zero function of the disorder strength characterized by $W$ and increases as the disorder strength is increased as well as when the superconducting gap becomes comparable to the nearest neighbor hopping. The function $\gamma_N(W)$ for other disorder distributions can be similarly extracted from the existing literature.

 {\it Oscillations.--- } While band oscillations are washed out, the wavefunction of the underlying Anderson problem can still have short range oscillations under the decaying envelope, but these are qualitatively different from the band oscillations in the uniform case. As noted in Subsection.\ref{sec:gen}, local variations of the Majorana wavefunctions, $a_n$, are governed by Heisenberg equations of motion, Eq.\ref{anderson}.
 
As emphasized throughout, this equation exactly corresponds to the Heisenberg equation of motion for the Anderson problem at zero energy. Solutions of the corresponding wavefunction have been avidly studied in past literature. The localized wavefunction typically exhibits random oscillations on the scale of the lattice spacing. They are heavily dependent on the disorder configuration and have large sample-to-sample fluctuations. They are known to have correlations on the scale of the mean free path and obey statistics independent of the decaying envelope ~\cite{Mirlin00, Ivanov12}. 

\begin{figure}[]
\subfloat[][]{\includegraphics[width=0.4\textwidth]{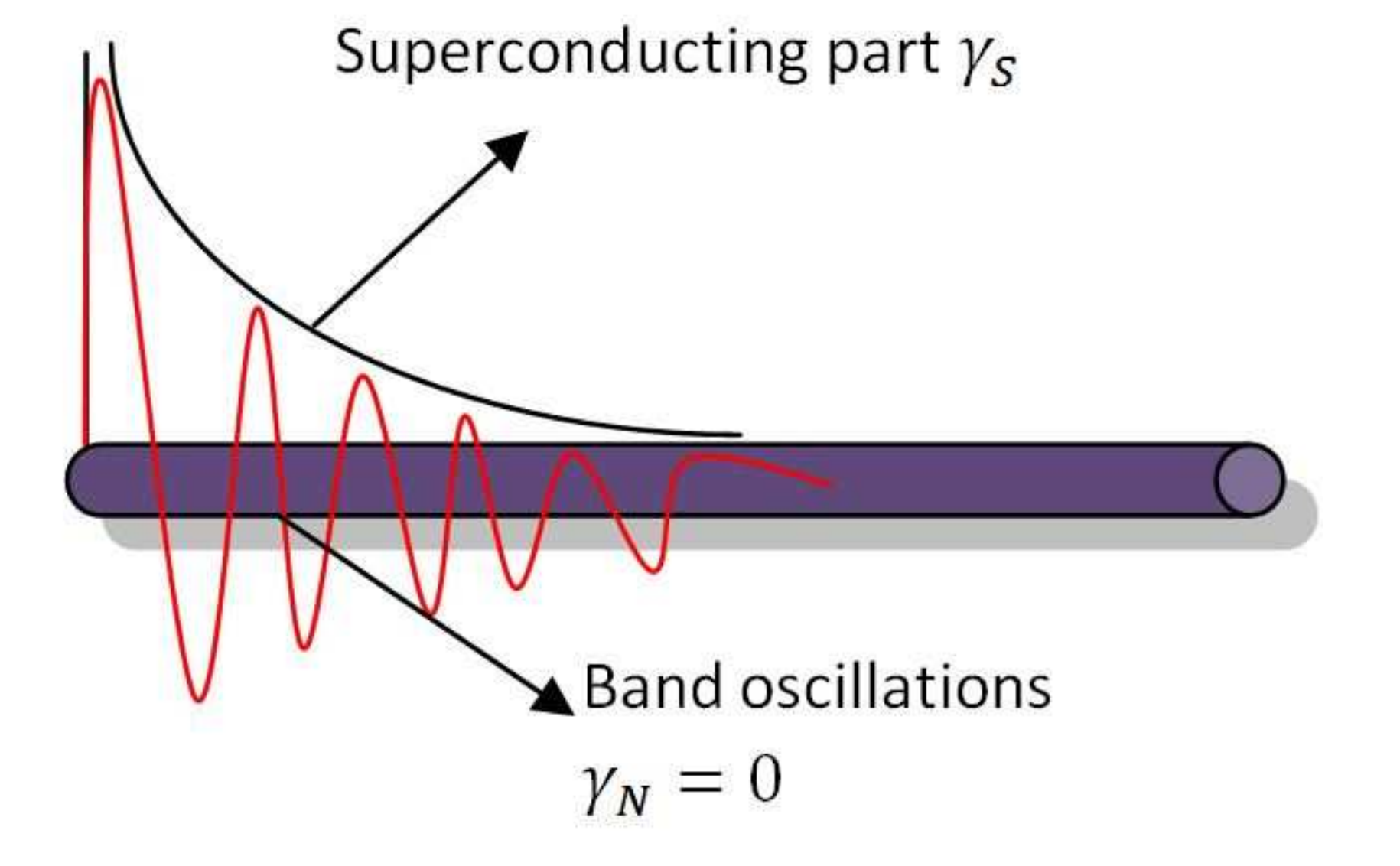}} \\
\subfloat[][]{\includegraphics[width=0.4\textwidth]{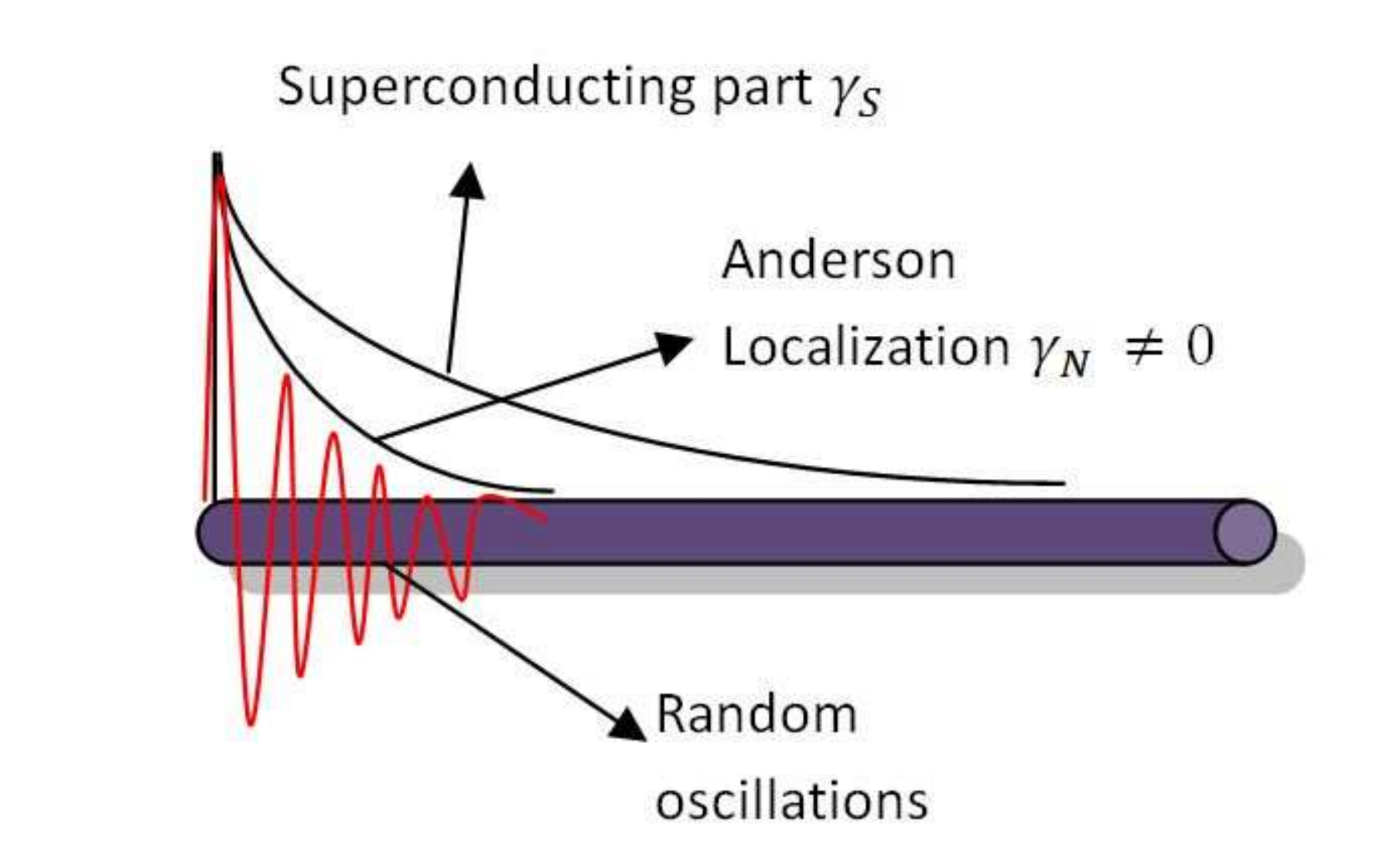}}
\caption {A schematic of the Majorana wavefunction in (a) the uniform case within the oscillatory regime and (b) the disorder case. (a) Within the parameter regime containing the circle of oscillations (see text), in addition to a decaying envelope having an associated Lyapunov exponent $\gamma_s$, stemming purely from the superconducting order, band oscillations are present. (b) For the disordered case, band oscillations are replaced by random oscillations and a second decaying scale associated with a Lyapunov exponent $\gamma_N$, both stemming from the underlying Anderson localization setup of a non-superconducting normal wire having the same disorder configuration.}
 \label{fig:disMajorana}
\end{figure}

  {\it Thus, disorder qualitatively changes the nature of Majorana wavefunction in that it i) imposes an extra localization scale in addition to the localization due to the superconducting gap and ii) changes the nature of underlying oscillations Fig.\ref{fig:disMajorana}.}

 This as we shall see has consequences on the finite-size splitting of ground state degeneracy and fermion-parity switches.

{\it Critical properties of Majorana wavefunction. --- }
Turning to the critical point separating the topological and non-topological phases, as discussed in previous sections, at this point, the Majorana decay length diverges. Equivalently, the Lyapunov exponent vanishes and thus, from Eq. \ref{xi_dis},  we have the relationship 
\beq
\gamma_S + \gamma_N = 0.
\eeq
For the specific case of Lorentzian disorder discussed above, this condition enables us to identify the critical point to be $W_c = 2\Delta$\cite{DeGottardi13A}. At this point, in going between the topological and non-topological states, the Majorana end modes completely extend into the bulk, in contrast to typical Anderson localization physics, and then vanish upon crossing the critical point. 

While we do not perform further analyses based on our transfer matrix techniques here, we invoke studies from previous work. Several insights on critical behavior in disordered Kitaev
 chains can be extracted by studying properties of the extensively studied spin-1/2 random transverse 
 field Ising chain. The Kitaev chain can be exactly mapped to a close relative, the transverse field XY spin chain ~\citep{Lieb61}. This XY model has two kinds of critical lines, an Ising type (ferro- to
  paramagnetic) and an anisotropic type (change in direction of magnetization). This transition from the ferro-
   to the paramagnetic phase corresponds to the the topological phase transition in the Kitaev chain of spinless fermions and is thus of interest in this context. The disordered Kitaev chain corresponds to
   the XY spin chain in the random transverse field and, in a particular limit ($\Delta = 1$), 
 to  the random field Ising model(RFIM), the critical properties of which are well known ~\cite{Fisher95,Balents97,Bunder99}. The ferro- to paramagnetic transition in this case belongs to the universality class of infinite disorder fixed point. One of the key features of the phase transition in RFIM\cite{Balents97,Fisher95} is the existence of two characteristic divergent length scales having different values of
  critical exponent $\nu$, where, as a function of distance to criticality, $\Delta$, each length 
  scale diverges as $\xi \sim \Delta^{-\nu}$. 
   One length scale, $\xi_{mean}$,
  characterizes the decay of average Green's function $C_{av}(r) \sim e^{-r/\xi_{av}}$, and
    diverges as $\nu_{av}=2$. The second, the typical localization length
     ,$\xi_{typ}$, reflects the  most probable correlation length. It can be extracted from the 
     single particle density-of-states and diverges with an exponent $\nu_{typ}=1$.
     
     As pointed out in Ref. ~\onlinecite{Gruzberg05}, the Lyapunov exponent of the transfer matrix at zero energy corresponding to the random matrix ensemble of Class D corresponds to the `typical' value of the correlation length. We expect the Lyapunov exponent of the Majorana transfer matrix to reflect the same properties and the correlation length, which in this case is the decay length of the Majorana, to have a critical exponent of $\nu=1$. In other words, the Lyapunov exponent corresponding to the Majorana transfer matrix vanishes near the topological phase transition to the trivial phase linearly as a function of distance to criticality. Further studies on the critical exponent of the 'mean` correlation length in the context of physics of Majorana modes are in order.

%

{\it Topological phase transition: Relating to result by Brouwer et al. ---}
 In addition to the observations made above, we can relate our work to results obtained in Ref.~\onlinecite{Brouwer11} on the disorder driven topological phase transition. 
 In this work, a condition is derived for the critical disorder strength for transition into non-topological phase as : $2\textit{l}=\xi $, where $\textit{l}$ is the mean free path in the disorder configuration and $\xi$ is the superconducting coherence length. This can be seen from our condition of cancellation of the components of LE at the transition : $|\gamma_S|=|\gamma_N|$. We have already identified $\gamma_S \sim 1/\xi \sim \Delta$. In order to relate to the result of \onlinecite{Brouwer11}, let us recall that $\gamma_N$ is the inverse localization length of the underlying Anderson problem. As known from previous literature Ref.\onlinecite{Thouless73}, the localization length is equal to the twice the mean free path: $\gamma_N = 1/(2 \textit{l})$. Thus our condition translates to : $1/\xi =1/(2\textit{l})$, which is precisely result of Ref. \onlinecite{Brouwer11}, obtained through calculations involving Langevin dynamics.

So far we have considered only semi-infinite wires to study the Majorana wavefunctions. In the next sections we consider finite sized wires to study fermion parity effects.
 
 \section{Fermion parity switches and mid-gap states in finite-size wires}
 \label{ParSwitch}
 \subsection{General discussion}
 \label{sec:ParityFlipGenDisc}
In the thermodynamic limit, the Kitaev chain in the topological phase has a doubly degenerate ground state. The two zero energy Majorana end modes corresponding to the degeneracy can be combined to form a non-local Dirac fermionic state. This non-local electronic state can either be occupied or unoccupied and this in turn determines the fermion parity of the entire many body ground state. Thus the double degeneracy also corresponds to degeneracy in fermion parity.


 For finite-sized systems, the degeneracy undergoes an exponentially small splitting $J$ due to the overlap of the Majorana wavefunctions. The associated parity states form an energy pair $\pm J$ closest to zero energy.  Now the fermion parity of the ground state is determined by the fermion parity of the lowest of these two states.  Explicitly, for two Majorana end modes depicted by $\Gamma_{R,L}$, the effective tunnel-coupled Hamiltonian is given by \cite{Kitaev01}
\begin{equation}
H_{eff} = iJ \Gamma_R \Gamma_L/2= J(\tilde{n}-1/2).
\end{equation}
 Here $\tilde{n}=\tilde{C}^{\dagger}\tilde{C}$  and $\tilde{C}=(\Gamma_L-\Gamma_R)/2$ is the non-local Dirac fermionic mode obtained from the linear combination of  Majorana end modes. The occupation $\tilde{n}=0,1$ determines the ground-state parity of the system. 
 In earlier sections, we discussed the regime in which the Majorana wavefunctions is endowed with an oscillatory component. Consequently, the tunneling amplitude of the two end Majorana modes too becomes  an oscillatory function of the parameters of the system, crossing zero at specific points in parameter space. Thus varying the parameters can lead to level crossings of these states and corresponding fermion parity switches. As explored in several recent works, these parity switches leave definitive signatures in various macroscopic phenomena such as the fractional Josephson effect, non-equilibrium quench dynamics and charge fluctuations in these systems\cite{Kwon04, Fu09, Hegde14, Ben-Shach15, Thakurathi14, Crepin14, Sau13, Beenakker13, Zazunov13}. Given the possibility of using Majorana modes as a tool for topological quantum computation, the $Z_2$ fermion parity has also drawn interest as a possible way of implementation of topological qubits \cite{Flensberg12, Hassler11, Bonderson11}. There have also been proposals for detecting parity effects of the Majorana zero modes in Josephson junctions using microwave spectroscopy \cite{Ginossar14, Vayrenyen15}.

An interesting feature of the uniform Kitaev chain is that the number of parity crossings as a function of parameters increases linearly with the system size ~\citep{Hegde14}. As a result, while the splitting between parity states varies exponentially, the ground state parity shows frequent parity switches in realistic systems and can have important effects. Here, we study these switches in depth for uniform as well as disordered systems. 
 
 The tunnel coupling between the Majorana modes at the ends is a good approximation for the splitting of the degenerate states and explaining the parity crossings. Below, we use the transfer matrix technique outlined in the previous sections to go beyond the approximation and obtain the precise points where the level crossings occur.

\subsubsection{ Pfaffian measure of fermion parity -}
The method of calculating the ground state fermion parity that we use in this work was formulated in Ref.~\onlinecite{Kitaev01} by A. Kitaev and is as follows. Consider the Majorana Hamiltonian of Eq. (\ref{majoranaH}) and the transformation $B$ that reduces the Hamiltonian to the canonical form, i.e., $D = B^T H_M B$. Here $D$ is an anti-symmetric matrix having non-zero matrix elements only along the first off-diagonal entries. It can be shown that the ground state parity of the system is related to the unitary properties of $B$.  Specifically, the parity of the system is given by
\beq \text{P(H)=sgn[det(B)]}. \label{detB} \eeq

  As a simple illustration of this expression for parity, its application to a two-site system is as follows. The Hamiltonian of Eq. \ref{majoranaH} in the Majorana basis $[a_1, b_1, a_2, b_2]$ for a two-site system is given by 
\[ \left( \begin{array}{cccc}
0                &  -i \mu/2       & 0              & i(-1+\Delta)/2 \\
i\mu/2           &  0              & i(1+\Delta)/2  & 0  \\
0                & -i(1+\Delta)/2  & 0              & -i\mu/2  \\
-i(-1+\Delta)/2  & 0               & i\mu/2         & 0\end{array} \right)\]
The Pfaffian for the above Hamiltonian is 
\begin{equation}
Pf(H)=\frac{\mu^2}{4}-\frac{1-\Delta^2}{4}
\end{equation}
Thus, the Pfaffian changes its sign at $\frac{\mu^2}{4}=\frac{1-\Delta^2}{4}$. This is for the specific case of $N=2$ and this result can be verified by explicitly diagonalizing the Hamiltonian and obtaining the matrix $B$. The general condition for the points where the Pfaffian changes sign for a Hamiltonian of N sites  is given in the next subsection.

\subsubsection{Majorana transfer matrix and parity crossings -}
 Here we describe how the Majorana transfer matrix can be used to track the occurrence of  zero energy crossings. In Sec.\ref{sec:Transfer} we presented the form of the Majorana transfer matrix corresponding to zero energy solutions confined to the ends of a wire. For finite sized systems, due to the hybridization of these two modes, in general, a zero energy solution does not exist and the corresponding transfer matrices couple the degrees of freedom associated with the two modes. For each transfer matrix to correspond to a strict zero energy solution, it must satisfy certain conditions imposed through the boundary conditions for Majoranas to be end bound states. These boundary conditions are as follows. 
 
 In previous sections, we showed that the set of individual transfer matrices $\tilde{A_n}$ in Eq. \ref{transnormal}  corresponds to that of the normal tight-binding problem, which carried all necessary information on the corresponding zero energy Majorana wavefunction in the presence of a finite superconducting gap $\Delta$. As the most general finite size situation applicable for any potential landscape, consider the transfer matrix relating the wavefunction at the first site to the wavefunction at the last site -
 \bea 
 \left( \begin{array}{c}
 \tilde{a}_{N+1} \\
\tilde{a}_N \end{array} \right) = \left( \begin{array}{cc}
\tilde{\mathcal{A}}_{11} & \tilde{\mathcal{A}}_{12} \\
\tilde{\mathcal{A}}_{21} & \tilde{\mathcal{A}}_{22} \end{array} \right) \left( \begin{array}{c}
\tilde{a}_1 \\
\tilde{a}_0 \end{array} \right),  \label{normalmatrix} 
 \eea
 The full-chain transfer matrix is given by $\tilde{\mathcal{A}}= \prod_n \tilde{A}_n$.
For the corresponding decoupled Majorana mode to exist, we demand that its wavefunction naturally be confined to the length of the wire. To impose this  boundary condition, we may introduce two fictitious sites at $0$ and $N+1$\citep{Hatsugai93}. As the condition for the existence of the mode, we then have
 \beq
 \tilde{a}_{N+1}=\tilde{a}_0=0
 \eeq
The transfer matrix gives the equations 
\beq
\tilde{a}_{N+1}= \tilde{\mathcal{A}}_{11}\tilde{a}_1 + \tilde{\mathcal{A}}_{12}\tilde{a}_0.
\eeq
The boundary conditions now gives a strict condition on the elements of the transfer matrix, namely
\beq
\tilde{\mathcal{A}}_{11}=0.
\label{A11}
\eeq
Thus for {\it any finite-sized chain}, the general condition for the existence of a zero energy solution is $\tilde{\mathcal{A}}_{11}=0$.

 For the case of a homogeneous chain, below we can explicitly illustrate how this condition is satisfied.

\subsection{Parity sectors in the uniform Kitaev chain}

\begin{figure}[]
\centering 
\subfloat[][]{\includegraphics[width=0.5\textwidth]{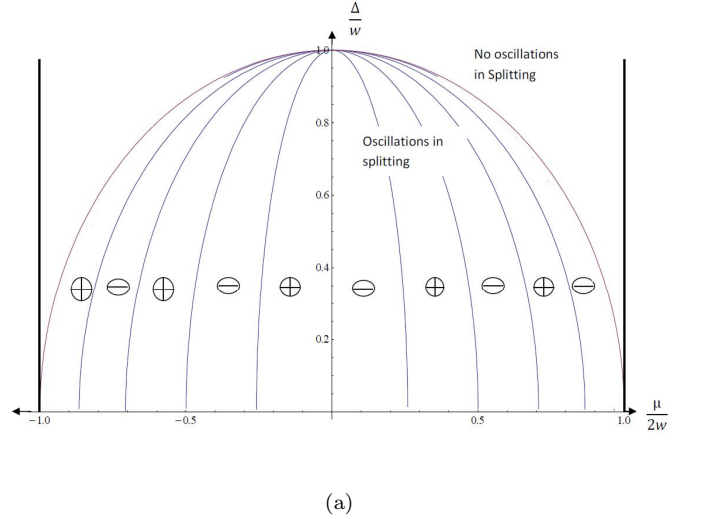}}\\
\subfloat[][]{\includegraphics[width=0.5\textwidth]{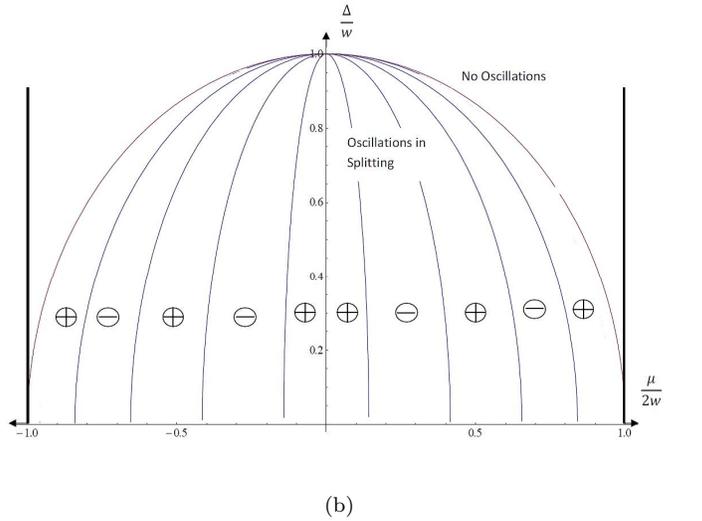}}
\caption {Ground state parity for a uniform {\it finite length} Kitaev chain within the topological phase in the phase diagram of Fig.\ref{fig:betacircle}. Within the circle bounding the region where Majorana bound state wavefunction exhibit oscillations, alternating parity sectors are demarcated by ellipses. These parity sectors depend on the length of the chain. For chains of odd length (N=11)(fig.(a)) the sectors are anti-symmetric across $\mu=0$ and are symmetric for chains of even length (N=10)(fig.(b)). Outside the circle, the Majorana modes are overdamped with no oscillations. }
 \label{fig:paritydiagram}
\end{figure}
The degeneracy splitting, level crossing and fermion parity switches can be tracked exactly in the case of a uniform chain. This was studied in our previous collaborative work in Ref.\onlinecite{Hegde14} and in other works \citep{Kao14, Zyavgin15}. To recapitulate, the finite-size degeneracy splitting introduces new features in the phase diagram of the Kitaev chain. The points at which the split-levels cross, thus restoring degeneracy even at finite size, form ellipses within the COO of the phase diagram (Fig.\ref{fig:paritydiagram}). The elliptical boundaries can be derived by enforcing this degeneracy condition on the transfer matrix of Eq.\ref{tmatrix} to yield
\beq
\Delta^2 +\frac{\mu^2 sec^2(\pi p/(N+1))}{4} = 1,
\label{ellipse}
\eeq

where $p=1,2...N/2$ for even $N$ and $p=1,2...(N-1)/2$ for odd N. These ellipses divide the circle into different parity sectors.  Consistent with Fig. \ref{fig:paritydiagram}, for fixed $\Delta$, parity crossings occur at chemical potential values satisfying
\beq
\mu_{switch}= 2\sqrt{1-\Delta^2} \cos\bigg(\frac{\pi p}{N+1}\bigg).
\eeq
Larger values of $p$ correspond to lower values of chemical potential.
 As each crossing is accompanied by a fermion parity switch, the adjacent areas across the elliptic boundaries are sectors of opposite parity. Thus for any given system size N, there are a number of parity sectors within the COO in the phase diagram.

{\it Even vs Odd number of sites. ---} It is important to note the difference in the features of the parity sectors for even and odd number of sites. For even number of sites, there is a symmetry in the parity sectors across the line $\mu=0$, whereas it is anti-symmetric for odd number of sites. This feature has a significant effect on  parity switches in the disordered case, as discussed later on.

{\it Majorana transfer matrix and parity switches. ---}
Here we show the manner in which the Majorana transfer matrix offers an effective means of tracking the fermion parity switches.
Consider the version of the individual transfer matrix of Eq.\ref{transnormal} that is applicable to the homogeneous case.  The chemical potential $\mu$ is the same on each site and the associated transfer matrix takes the form 
 \bea 
 \tilde{A}_n &=& \left( \begin{array}{cc}
-\frac{\mu}{\sqrt{1-\Delta^2}} & - 1 \\
1 & 0 \end{array} \right) \eea
 Its eigenvalues are given by $\lambda_{\pm}= e^{\pm i \alpha}$ where $\alpha=\tan^{-1}\bigg(\frac{\sqrt{4(1-\Delta^2)-\mu^2}}{\mu}\bigg)$. 
 Using Eq. \ref{ellipse} and as elaborated in Ref.~\onlinecite{Hegde14}, we have the condition on $\alpha$ for the zero-energy crossings 
 \beq
 \alpha =\frac{\pi p}{N+1}
 \eeq
 $p=1,2...N/2$ for even N and $p=1,2...(N-1)/2$ for N odd. 
  The full-chain transfer matrix can be calculated exactly using Chebyshev's identity for uni-modular matrices ~\citep{Markos} to yield
\beq
\left( \begin{array}{cc}
-\frac{\mu}{\sqrt{1-\Delta^2}} & - 1 \\
1 & 0 \end{array} \right)^N  =  
\left( \begin{array}{cc}
-\frac{\mu}{\sqrt{1-\Delta^2}}U_{N-1}-U_{N-2} & \: U_{N-1} \\
U_{N-1} & \: -U_{N-2} \end{array} \right)
\label{Chebyshev}
 \eeq
where $U_N = \sin(\alpha(N+1))/\sin\alpha$. Now the condition for the existence of an edge state reads
\beq
[\tilde{A}^N]_{11}= -\frac{\mu_n}{\sqrt{1-\Delta^2}}U_{N-1}-U_{N-2}=U_N =0
\eeq
 Here we have used the identity $2\cos\alpha U_{N-1} - U_{N-2}= U_N$. Therefore the condition for zero-energy crossings becomes
 \beq
 \sin(\alpha(N+1))/\sin\alpha =0
 \eeq
which is in fact satisfied precisely when $\alpha = p \pi/(N+1)$.

\begin{figure}[]
\centering 
\subfloat[][]{\includegraphics[width=0.4\textwidth]{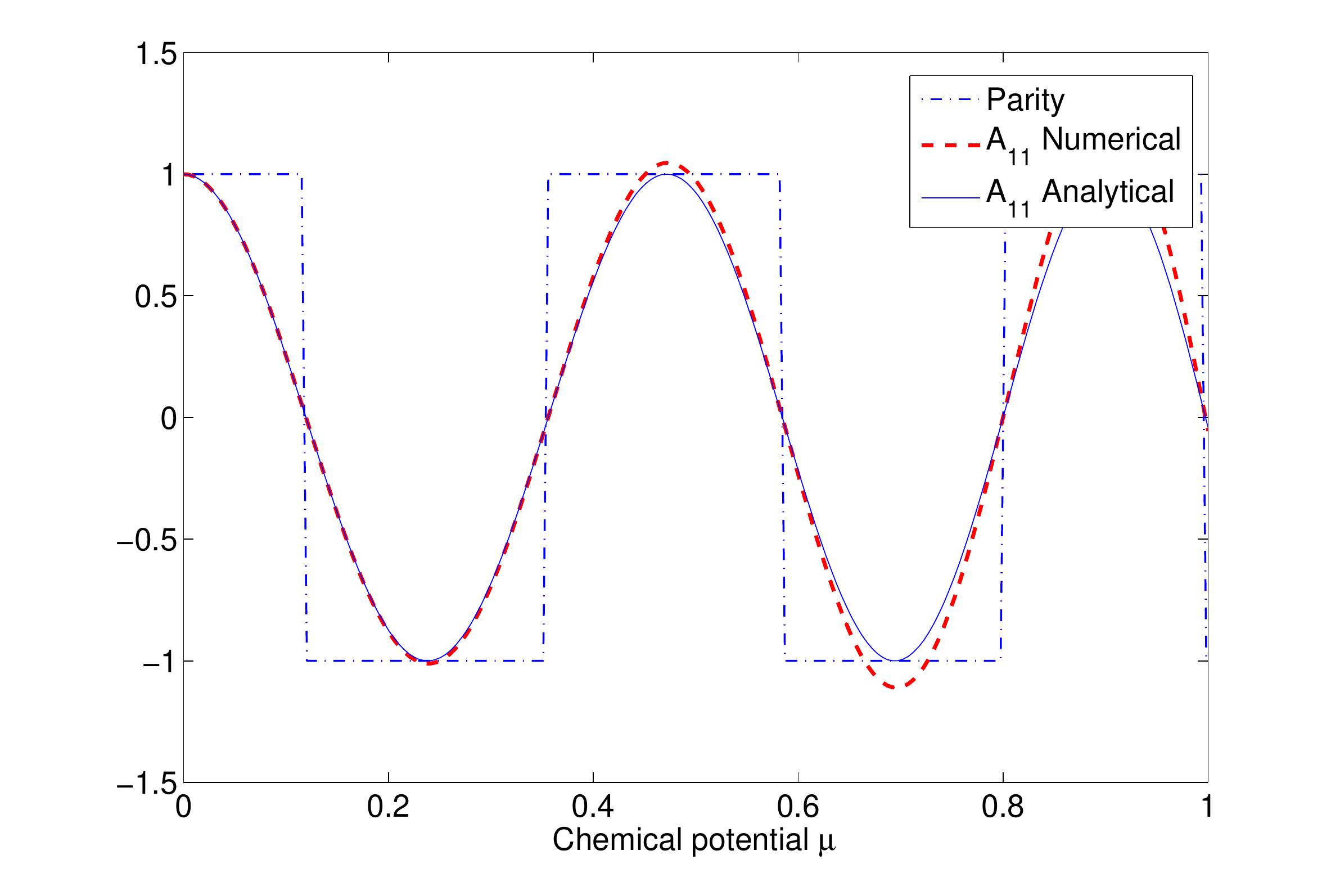}}
\caption {Plots which show the comparison between the fermion parity in a uniform Kitaev chain,  calculated using Eq.\ref{detB}, and the matrix element $\tilde{\mathcal{A}}_{11}$ of the zero-energy Majorana transfer matrix whose vanishing value reflects the existence of a zero-energy state. The matrix element is calculated both analytically using Eq.\ref{Chebyshev} and numerically for a uniform chain. One can see that the parity switches coincide exactly with the matrix element going to zero. Here $N=21,\Delta=0.6$}
 \label{fig:betaparity}
\end{figure}

 From the above, we find that at each level crossing, $[\tilde{A}^N]_{11}$ tends to $0^{\pm}$ depending on $p$ being an even/odd value. 
  Figure \ref{fig:betaparity} shows the numerical results for parity switches and the sign of $[\tilde{A}^N]_{11}$ identically track each other, confirming our arguments.

\subsection{Scaling of parity swtiches in the homogeneous chain}
Parity switches as a function of the chemical potential in the homogeneous chain exhibit an interesting scaling behavior. These parity switches are found only within the COO of the topological phase in which the Majorana wavefunctions are oscillatory. Within this regime, if we consider the parity switches for the chains with same length but different values superconducting gaps, the variation of the chemical potential in each case can be scaled to collapse all switches to a single curve. One can understand the collapse as follows: start with a system having a zero superconducting gap and obtain the switches in parity as function of $\mu$. Using this, one can reproduce the parity switches at any value of finite gap $\Delta$ by scaling $\mu$ by a  $\Delta$-dependent factor, which as seen in Eq.\ref{anderson}, is $\sqrt{1-\Delta^2}$.

\begin{figure}[]
\centering 
\subfloat[][]{\includegraphics[width=0.4\textwidth]{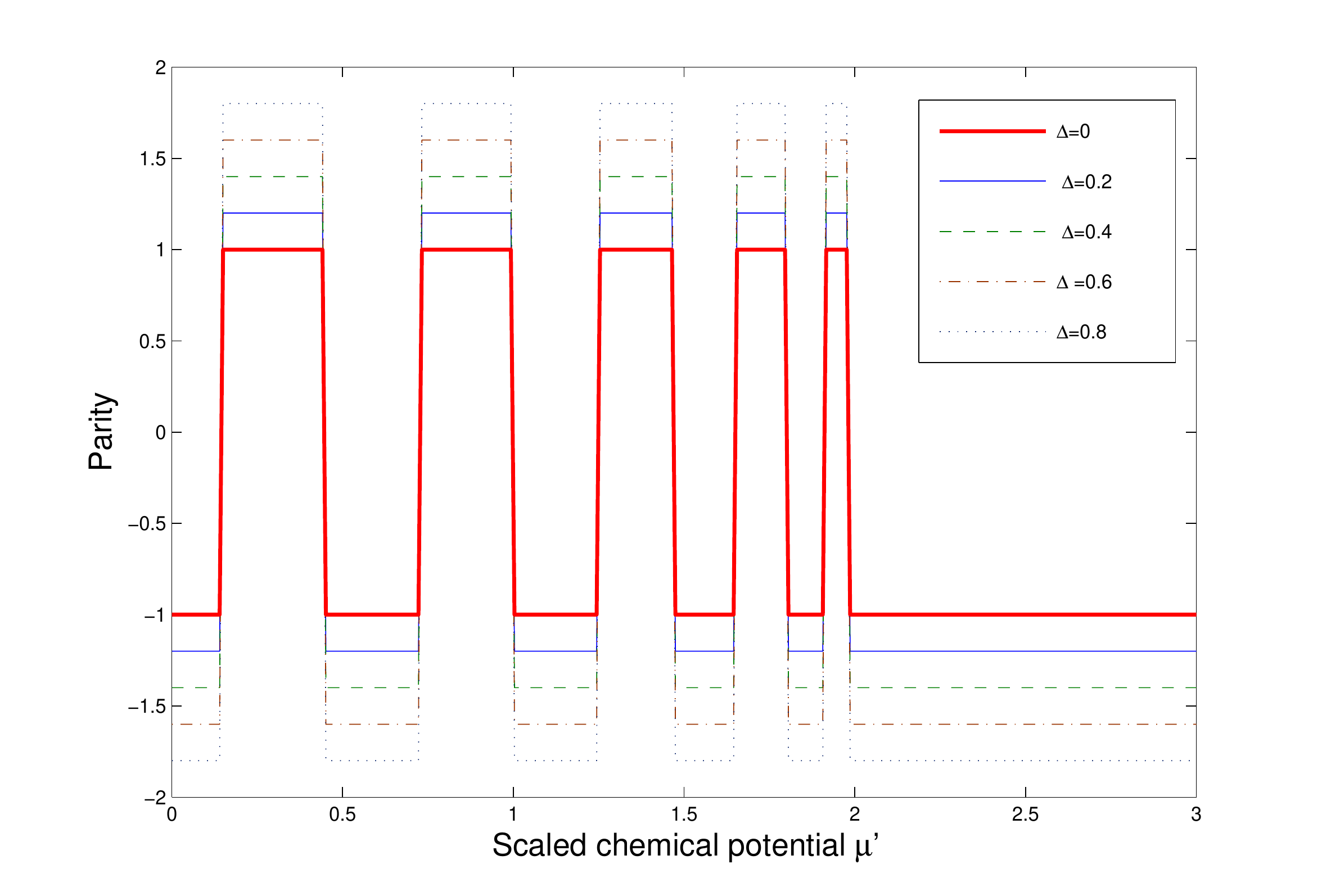}}
\caption {Fermion parity switches are concurrent in wires of differing superconducting gap $\Delta$ as a function of the scaled chemical potential $\mu'=\mu \sqrt{1-\Delta^2}$. This is shown here for the uniform Kitaev chain of length $N=20$, where parity is calculated using Eq.[\ref{detB}]. The thick red plot is for $\Delta=0$. The other plots (scaled away from unity for proper visibility) are for finite superconducting gaps.}
 \label{fig:scaling}
\end{figure}  

Fig.[\ref{fig:scaling}] shows the collapse of the scaled parity oscillations at different superconducting gaps for the homogeneous case. The collapse of all the plots appropriate for different values of the superconducting gap $\Delta_i$ is expressed as

\bea 
P(H[\mu, \Delta=0,L])=P(H[\mu \sqrt{1-\Delta_i^2},\Delta_i,L]) 
\label{parityscale}
\eea
This scaling within the COO of the topological phase can be understood from the previous arguments on the origin of the oscillations from an underlying tight-binding model in the absence of a gap. The oscillations of Majorana wavefunctions are given by the solutions of the equation Eq. ~\ref{normaloscillations}, which is the Heisenberg equation of motion for the normal system having the scaled chemical potential. For $\Delta=0$, the scaling factor $\sqrt{1-\Delta^2}$ is 1 and the wavefunction oscillations are functions of only $\mu$. For finite gap $\Delta$, the wavefunction oscillations are functions of a scaled down chemical potential $\mu/\sqrt{1-\Delta^2}$. Therefore, the resulting parity oscillations for the gapless case can be visualized to be `stretched out' along the axis of $\mu$, when compared to the case of finite gap. To map the parity oscillations in the gapped case to the gapless case,we need to scale up $\mu$ to $\mu\sqrt{1-\Delta^2}$ in the gapped Hamiltonian, such that it cancels the scaling factor of $\mu$ in the underlying gapless, normal tight-binding model. This is shown in Eq.[~\ref{parityscale}].

In fact, this scaling can be obtained as function of length of the chain too. Such a `universal scaling' has been reported in the context of entanglement spectrum of the transverse field XY spin chain ~\citep{Giampolo09}. This spin chain has an exact mapping to the Kitaev chain. From the correspondence between the spin chain entanglement spectrum and edge spectrum in topological superconductors\citep{Fidkowski10},it is natural that there be a mapping between the Schmidt gap oscillations in the spin chain and parity oscillations in the Kitaev chain.

\section{Fermion parity switches and low-energy states in finite-sized disordered wires}
\label{DisParity}
\subsection{Energy spectrum and Majorana energy splitting}
Disordered Kitaev chains have formed the topic of active research for over a decade for several reasons\citep{Motrunich01, Brouwer00,Brouwer03,Gruzberg05,Reider14, Reider13, Reider12, Pientka12, Brouwer11, Neven13, Bagrets12, Hui14, Lobos12, Fregoso13, Stanescu12, Sau13A, Beenakker14, Pikulin12, Fulga11, Akhmerov11, Beenakker13, DeGottardi13A}. Belonging to the symmetry classes D and BDI (depending on whether they respect time reversal symmetry breaking or not), these systems exhibit behavior that starkly deviates from that of their normal counterparts. One of the highlighting features is the presence of a delocalization-localization transition as a function of disorder strength\citep{Motrunich01, Brouwer00,Brouwer03}.  This transition also corresponds to the transition between the topological phase and non-topological phase, as shown in Ref.\onlinecite{DeGottardi13}. In contrast to Anderson localization physics of normal systems in one-dimension, the critical point in fact forms a mobility edge separating two localized phases. The mobility edge itself possesses a zero energy multifractal state that extends through the bulk of the entire system and offers a route for the Majorana end modes in the topological phase to permeate and disappear into the bulk upon entering the non-topological phase\citep{Motrunich01, Brouwer03}. Here, building upon known salient features of disordered Kitaev wires, we focus on finite sized systems and the behavior of the zero-energy degeneracy splitting, associated Majorana wavefunction physics, and parity switches. We use the `box disorder' for numerical studies, where the values of the chemical potential $\mu_n$ is taken from a uniform distribution of width W. Such a distribution with zero mean is given by
\beq
\mu_n = \bigg[-\frac{W}{2},\frac{W}{2} \bigg]
\label{boxdis}
\eeq

\begin{figure}[]
\subfloat[][]{\includegraphics[width=0.4\textwidth]{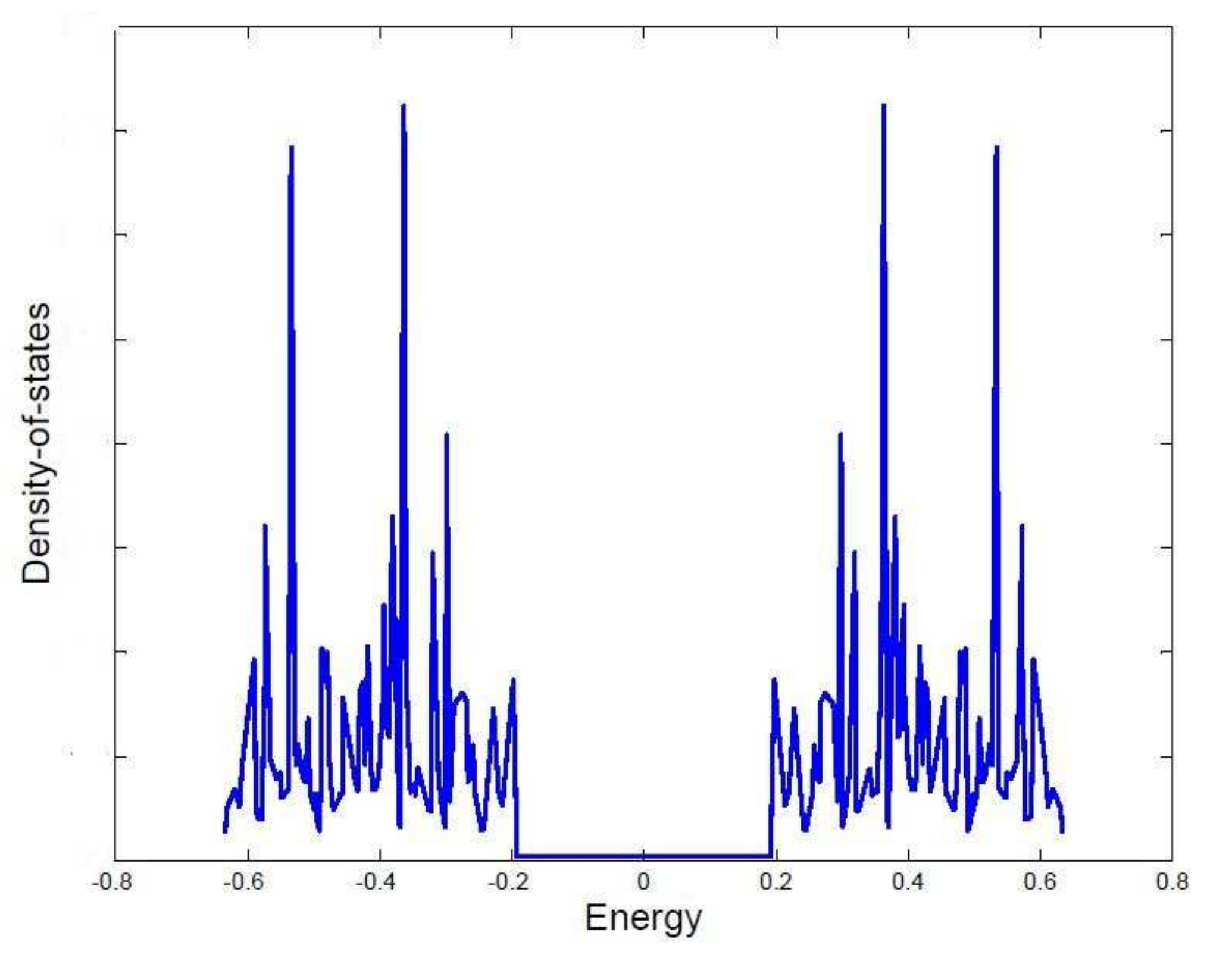}}\\
\subfloat[][]{\includegraphics[width=0.4\textwidth]{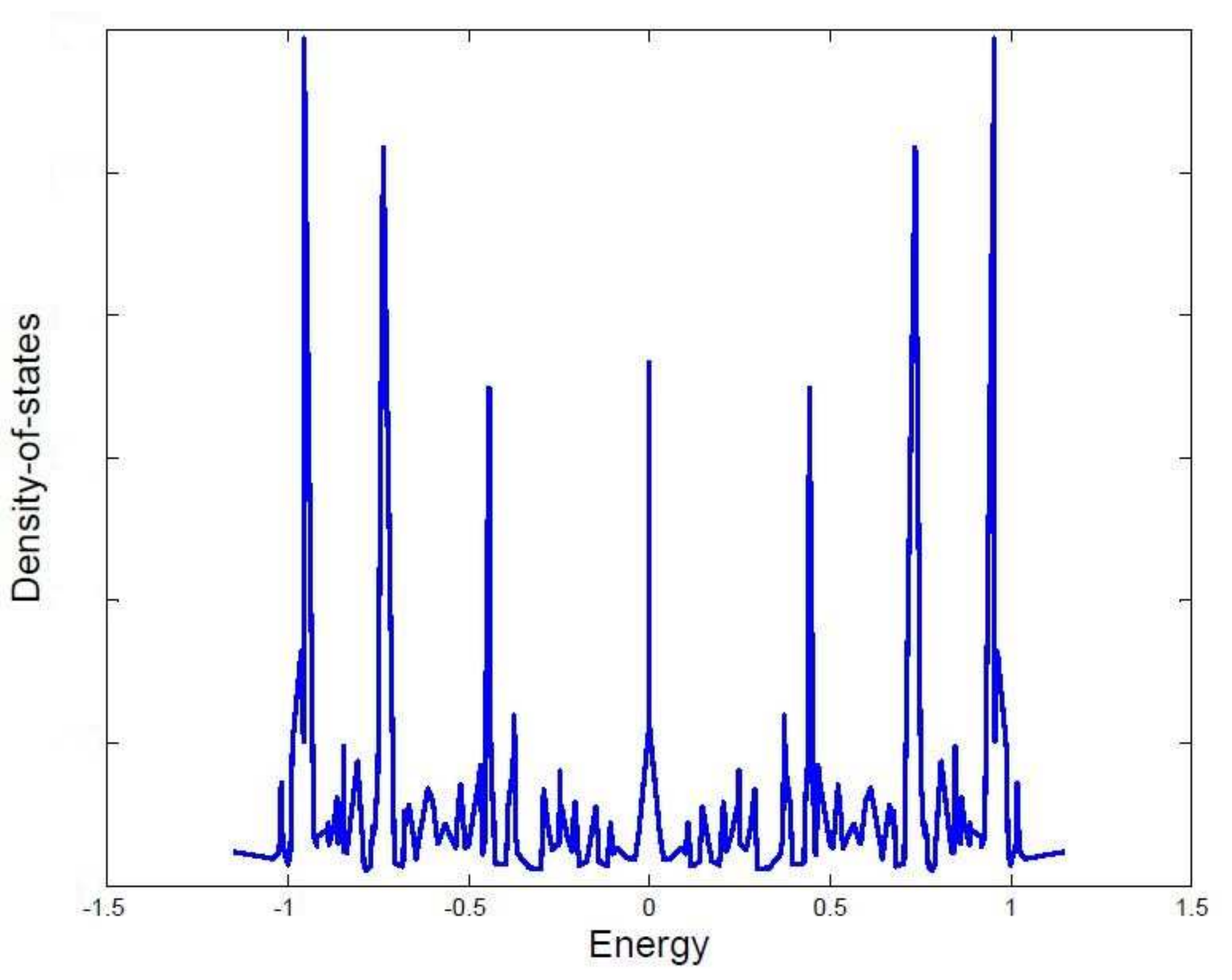}}\\
\subfloat[][]{\includegraphics[width=0.4\textwidth]{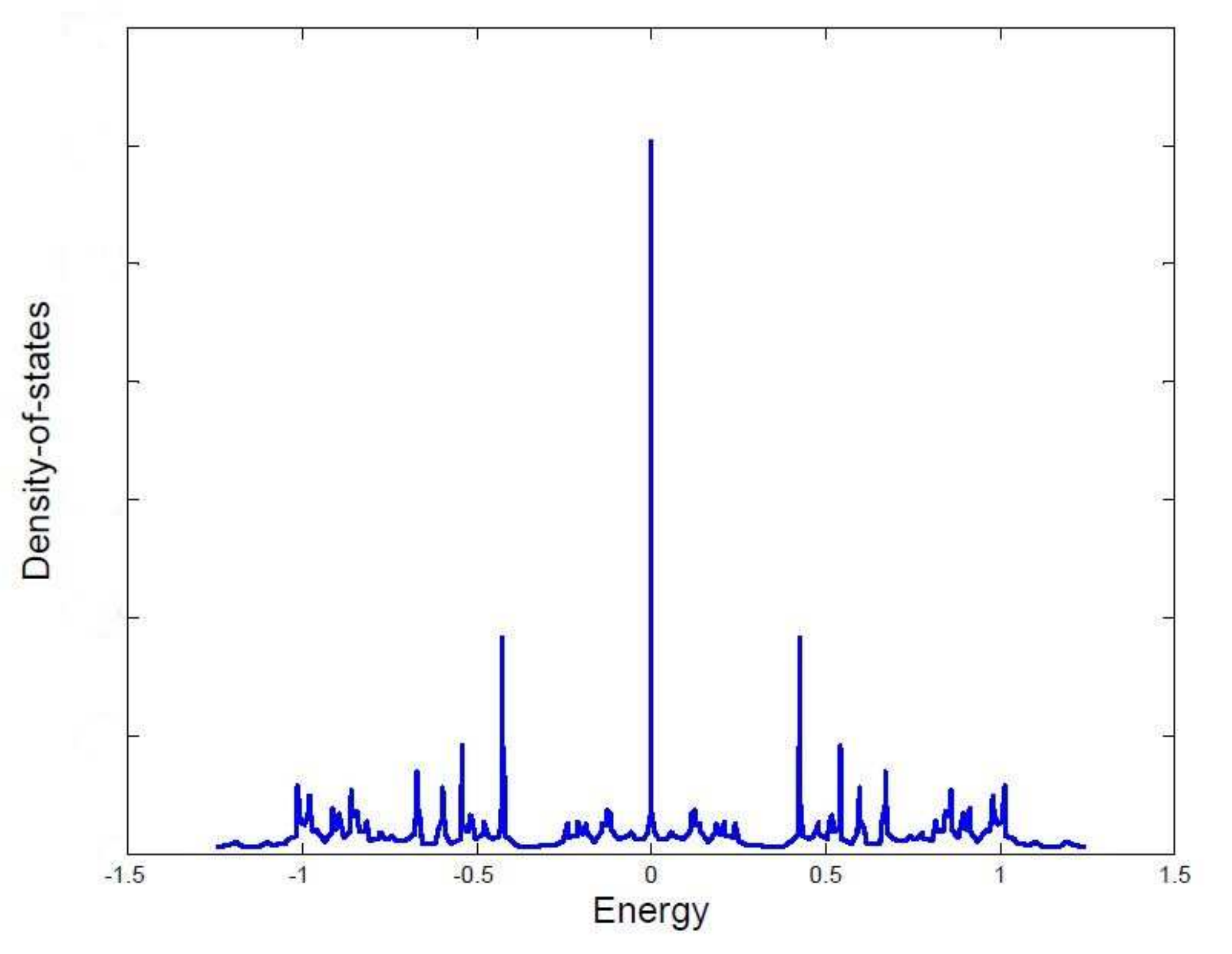}}

\caption {Density-of-states  plots for a disordered Kitaev chain ($\Delta=0.6$) as a function of distance from the Fermi energy for a single disorder configuration of box disorder \ref{boxdis}. (a) For weak disorder ($W< 8$), there is a well-defined superconducting gap in the density-of-states. (b) As the disorder strength is increased, the gap is filled due to the proliferation of low-energy  bulk states.(c) At a critical disorder strength, there is  `singularity' in the density-of-states. The behavior beyond the critical point resembles that of (b).}
 \label{fig:disorderdos}
\end{figure}

{\it Evolution of the density-of-states (d.o.s.). ---} 
A striking feature of the disordered Kitaev chain is the presence of a large gapless regime around the phase transition. As a function of disorder strength, low disorder smears the d.o.s near the gap edges, filling in some previously forbidden states. Increasing the disorder results in a proliferation of low energy localized bulk states, followed by a divergent d.o.s about zero energy, corresponding to a Griffiths-like phase. Further increase in disorder results in a divergent d.o.s that respects universal dependence on energy $\epsilon$ of the form $\epsilon^{-1}|\epsilon|^{-3}$ at the critical point, corresponding to the Dyson singularity\cite{Motrunich01,Gruzberg05}. Still further, the system enters the non-topological gapless phase.  Several investigations on the distribution of these mid-gap states have explored scaling of density-of-states in Griffiths phases, transport properties  and topological phase transitions in the context of disordered Majorana wires. ~\cite{Motrunich01, Reider14, Reider13, Reider12, Pientka12, Brouwer11, Neven13, Bagrets12, Hui14, Lobos12, Fregoso13, Stanescu12, Sau13A, Beenakker14, Pikulin12, Fulga11, Akhmerov11, Beenakker13, DeGottardi13A}. Here, as a simple illustration, in Fig. \ref{fig:disorderdos}, we show the evolution of the d.o.s  for a single sample as a function of increasing disorder strength using exact numerical diagonalization. While detailed features cannot be resolved through our methods, the evolution clearly demonstrates the trends described above.

{\it Behavior of low-energy mid-gap states and Majorana physics. --- } Reflecting the behavior of the d.o.s, at very low disorder, the only low energy states correspond to the Majorana end modes. In finite sized wires, these states hybridize and exhibit a zero-energy splitting. Upon increasing disorder, while low-energy bulk states proliferate into the gap, the lowest energy states in the topological phase still correspond to robust Majorana modes (typically hybridized). Extensive applications of random matrix theory (RMT) for class D in previous work [\onlinecite{Beenakker14, Beenakker13} and references therein] show that the spacing between energy levels of the disordered Kitaev chain  (class D) respect the probability distribution of energy levels given by \citep{Altland97}]


\bea
P({E})d{E}= \prod_{i<j} |E^2_i-E^2_j|^{\beta} \prod_k |E_k|^{\alpha} e^{-E_k^2/v^2}dE_k  \\
\text{Class D :}  \quad \alpha=0  \quad \beta=2 \non
\label{RMTform}
\eea

Here $v$ is the variance of the distribution.
The exponent $\beta$, which measures level correlations between energy states that are not particle-hole symmetric, is non-zero.
Thus, for any two such energy states, level repulsion ensures that the probability of the two energy levels crossing goes to zero, avoiding any crossings. For two states with energies $\pm E$, however, since $\alpha=0$, the level crossing is allowed.

  \begin{figure}[]
 \subfloat[][]{\includegraphics[width=0.4\textwidth]{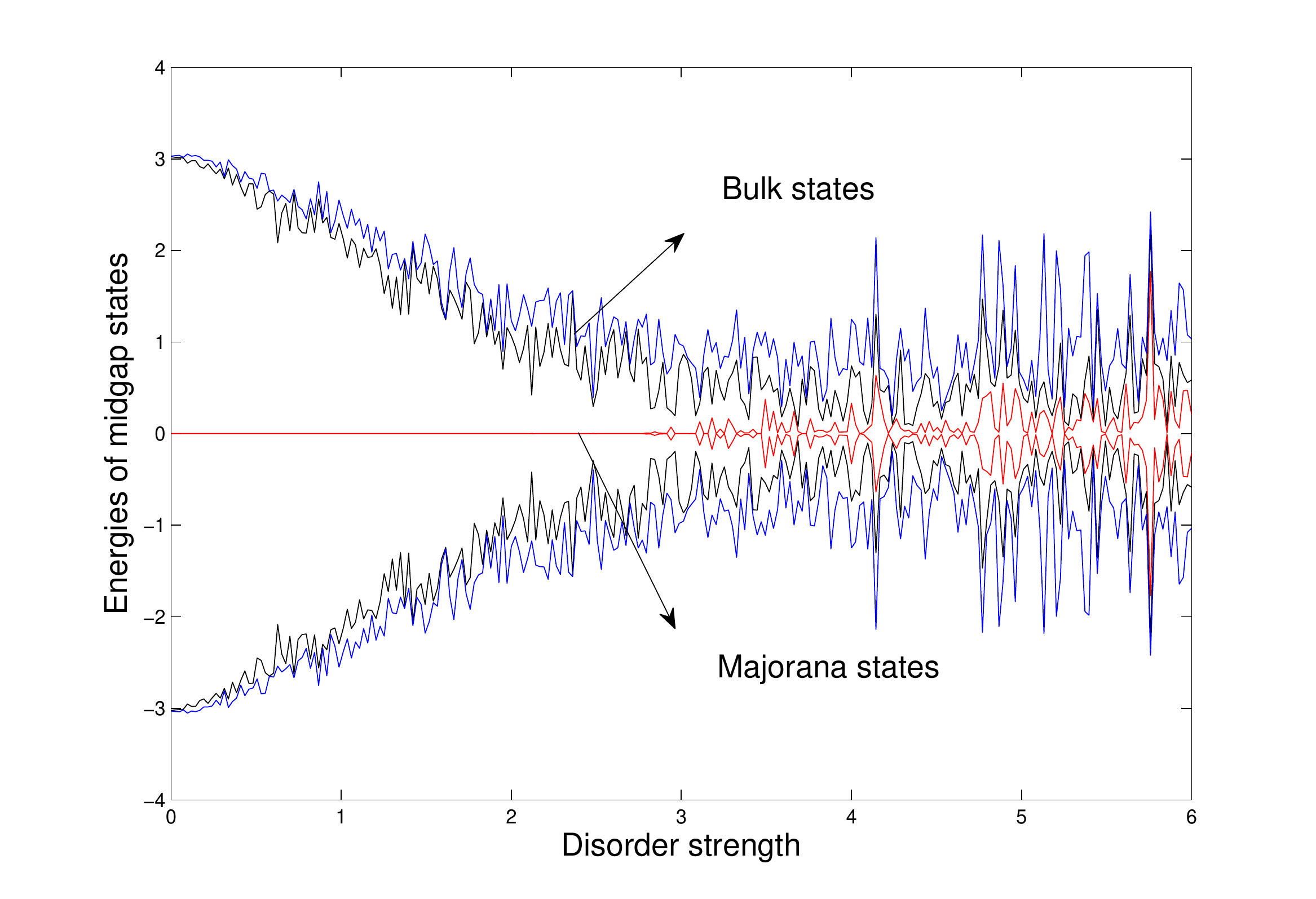}}\\
 \subfloat[][]{\includegraphics[width=0.4\textwidth]{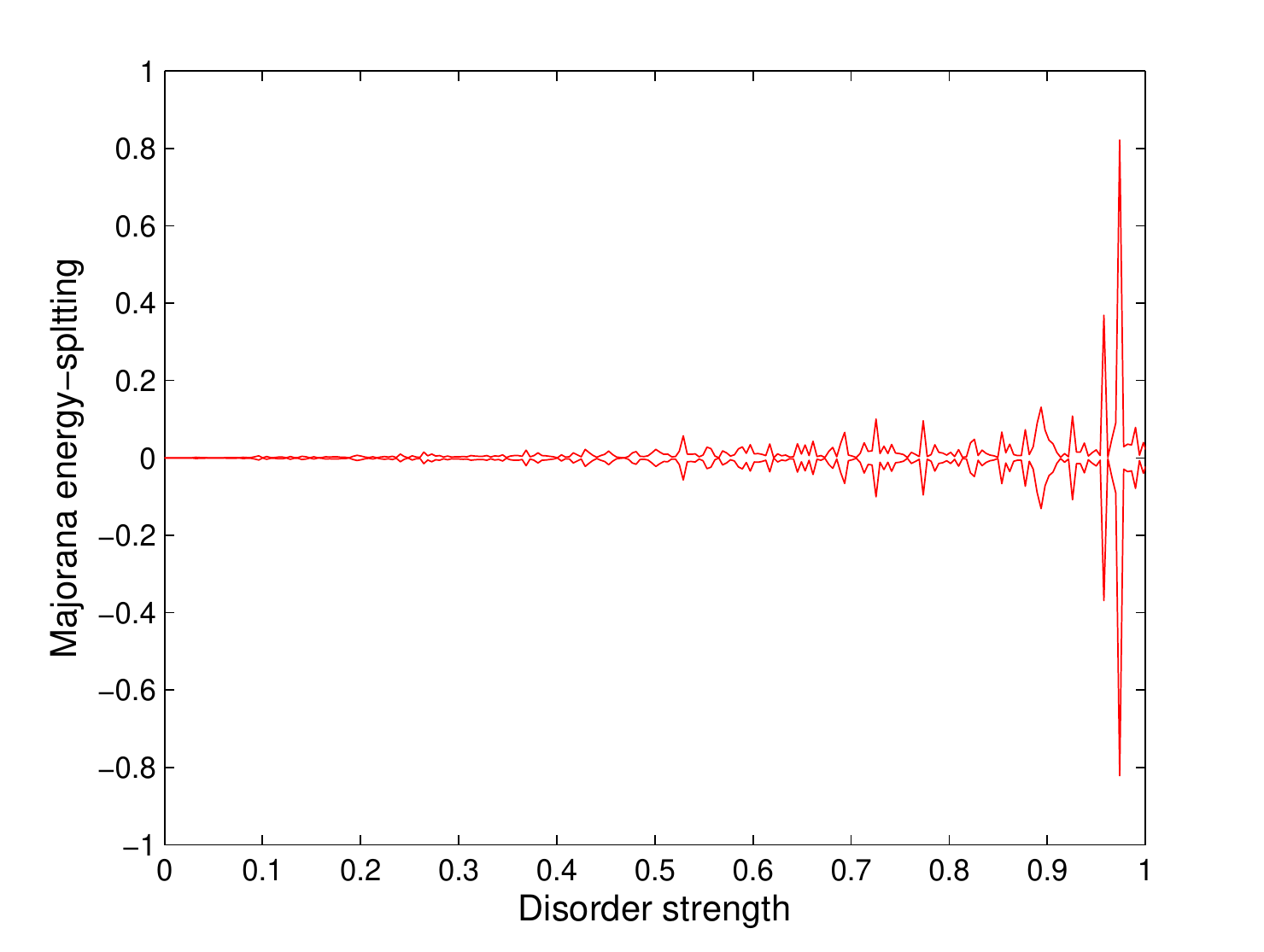}}\\
 \subfloat[][]{\includegraphics[width=0.4\textwidth]{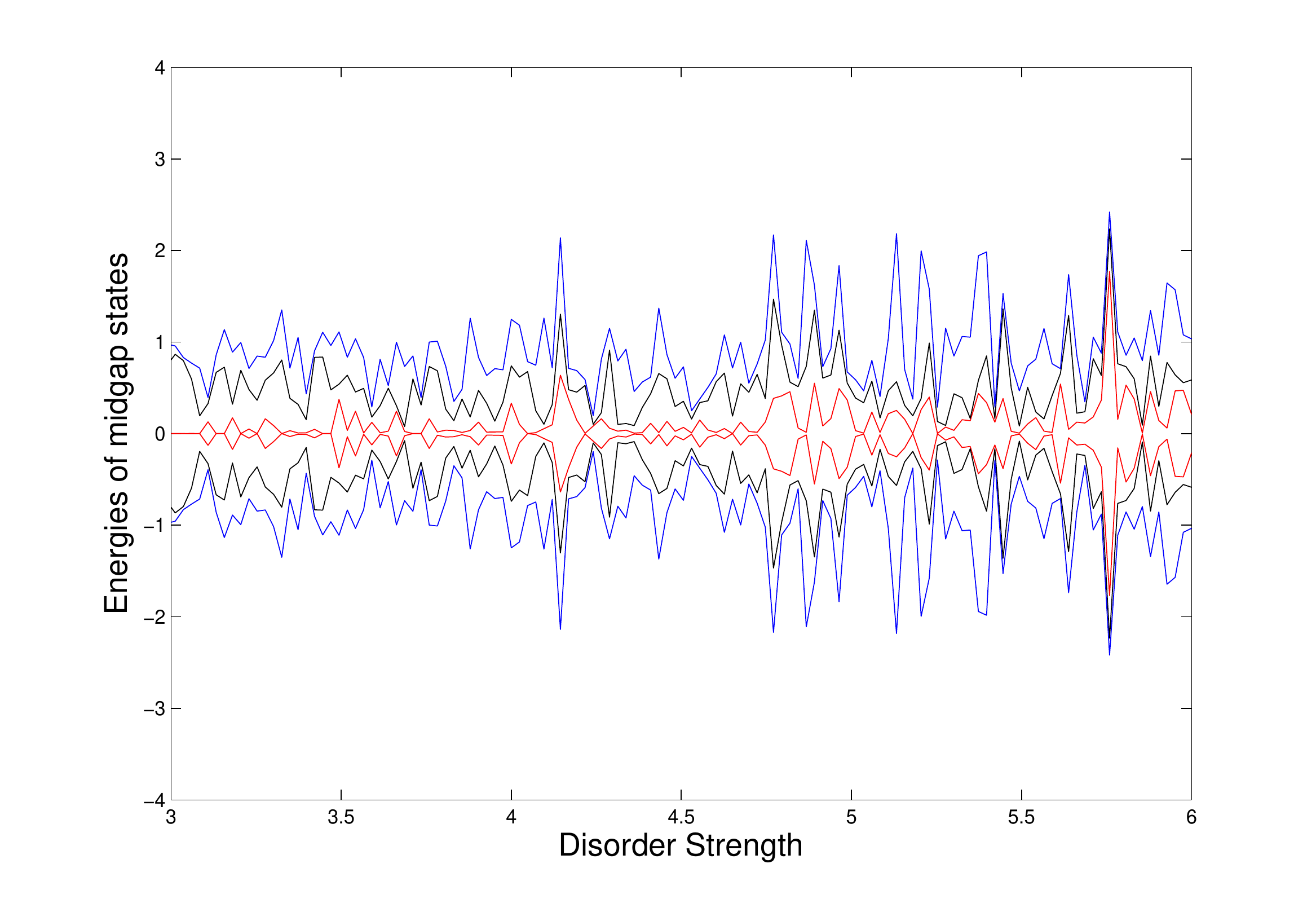}}
\caption {(a)Variation of a set of lowest energy levels (first 3 states) of the Kitaev chain as a function of disorder strength for box disorder \ref{boxdis} and other parameters fixed to $\Delta=0.6$, $N=30$. For small disorder width, the Majorana states are well separated from the bulk by a superconducting gap. As the disorder is increased there is a proliferation of the bulk states into the gap.(b) Zoomed in view of the two Majorana states split due to finite size. Their scale is exponentially suppressed compared to the bulk states. These states cross zero energy as the disorder width is varied, inducing a fermion parity switch in the ground state. (c) Griffiths phase: At strong disorder, there is an accumulation of large number of bulk states near zero energy. Level crossings between these states are forbidden due to level statistics of Class D (level repulsion). Nearing the critical disorder strength, the magnitude of the energy states due to Majorana splitting become comparable to the bulk states.}
 \label{fig:disorderstates}
\end{figure}

In Fig.\ref{fig:disorderstates}, we show the evolution of the behavior of the lowest three particle-hole symmetric energy pairs as a function of disorder. As described above, for low disorder, only two energy levels corresponding to Majorana modes can be seen close to zero energy. As the disorder strength is increased, other levels converge towards zero energy in close succession. Disorder causes fluctuations in the energy level spacing. For a significant portion of the topological phase, the scale of energy level splitting is very different for the Majorana modes compared to the next higher energy levels. As in the uniform case, the splitting of the Majorana modes as a function of system size is expected to be exponentially small while that of the other states is expected to be algebraic \cite{Brouwer11}. It can be seen that beyond the critical disorder strength for the topological phase to exist, the lowest energy modes now lose their Majorana character and their level splitting is comparable to that of the other modes.

Focusing on the energy level splitting of the Majorana modes, the behavior of the disordered Majorana wavefunction discussed in Sec.\ref{sec:wavefun_disorder} ought to dictate the splitting. We saw that three components characterize the wavefunctions:  i) the gap-protected robust envelope, whose localization is determined by the magnitude of the superconducting gap ii) a second decaying envelope due to disorder, whose localization is determined from the underlying Anderson problem (see Eq.\ref{xi_dis}) iii) and the sub-envelope  random oscillations dictated by the same Anderson problem. The first two determine the average scale of energy splitting of the degenerate zero-energy states in a finite size wire. The crossing of these states is determined by the third aspect; in the next subsection, we study these crossings in detail and their direct connection with fermion parity switches.

\subsubsection{Scaling of degeneracy-split states}

In Ref. \onlinecite{Brouwer11}, it was shown using a scattering matrix approach that the energy level splitting of coupled Majorana end  states in a finite sized wire is exponentially small compared to bulk states in the weak disorder limit. Due to large fluctuations, the energies of these mid-gap states themselves do not follow a simple probability distribution. But as is common with random systems, the logarithm of these energies obeys a normal distribution. A central result of Ref. \onlinecite{Brouwer11} is that for a continuum model of the p-wave superconducting wire, the average of this quantity has the form
\beq
  \langle \text{ln}(\epsilon_{0,max}/{2\Delta})\rangle =-L[1/\xi - 1/(2\textit{l})] ,
  \label{BrouwerEq}
\eeq
  where $\epsilon_0$ is the Majorana end state energy in a finite sized wire,$\Delta$ is as usual the magnitude of the superconducting gap,  $\xi$ is the superconducting coherence length and $\textit{l}$ is the mean free path of the corresponding disorder configuration.

These results can be understood in the light of our discussion on Majorana transfer matrices. As described before, the degeneracy-split end modes have energies proportional to the overlap of the Majorana wavefunctions. 
In a finite sized wire of length L, for two end modes having a decaying envelope of localization length $1/\gamma$, this overlap is proportional to $\epsilon_0 \sim e^{\gamma x}e^{\gamma(L-x)} = e^{\gamma L}$. Here, $\gamma$ is the Lyapunov exponent and is a negative quantity in the topological phase. As shown in Eq. \ref{LE2},  $\gamma L= (\gamma_S + \gamma_N)L$ consists of a superconducting and normal piece. 
Now, as is commonly invoked in treatments of localization physics and random systems, given the multiplicative nature of transfer matrices, the Lyapunov exponent corresponding to the transfer matrix is in general self-averaging. Thus, we expect  $\text{ln}(\epsilon_0)$ to have an average value of $L \gamma$; this result is reminiscent of Eq.\ref{BrouwerEq} where $\xi$ is the length scale associated with superconductivity and $\textit{l}$ with normal localization properties.

  
\subsection{Parity switches - qualitative discussion and numerical results}
In Sec.\ref{sec:ParityFlipGenDisc}, we outlined how a pair of zero energy Majorana modes form a Dirac fermion state that can be occupied or unoccupied, corresponding to two states of opposite parity. We then showed the manner in which end Majorana modes hybridize in a finite sized uniform chain, giving rise to an energy splitting and an associated unique ground state parity. We charted out the points in phase space where zero energy crossings take place, corresponding to ground state parity switches, and mapped the regions of odd and even parity in the topological phase diagram. Turning to disordered systems, in the previous section, we gave a detailed description of the behavior of low energy states and discussed the distribution of energies associated with zero energy splittings. Here, we focus on the vanishing of this splitting and show that this occurrence directly corresponds to a ground state parity switch. We also show the manner in which our study of the uniform system informs the ground state parity distribution map for the disordered case.  

{\it Zero-energy crossings and parity switches. ---}
Following the discussion on mid-gap states in the previous section, first off, we see that zero-energy level crossings are infact possible as a function of system parameters, such as chemical potential and disorder strength. The RMT result of Eq.\ref{RMTform} further corroborates that particle-hole symmetric states can undergo zero-energy crossings but other pairs of adjacent states cannot do so due to level repulsion. This suggests that the only states under-going zero-energy crossing are the ones associated with Majorana physics. In other words, zero-energy crossing are concurrent with fermion parity switches. We now demonstrate this explicitly. 

In Sec.\ref{sec:wavefun_disorder}, we saw that the Majorana wavefunction in the disordered case has a decaying enveloping as well as oscillations that are completely dictated by the underlying normal Anderson tight-binding model. In the previous section, we saw that the decaying envelope is directly related to the scale of the average zero-energy splitting and is always finite for a finite length wire. However, the oscillations directly contribute to the fluctuations and, in particular, to the vanishing of the splitting. In principle, just as the correlation between two decaying envelopes gives the average scale for zero-energy splitting, analytic studies of correlations between the oscillations\citep{Ivanov12} ought to give precise information on the locations where the splitting vanishes. Here, we resort to the numerical methods that we employed in previous sections. Specifically, we use the normal system transfer matrix condition for the existence of a zero energy state given by Eq.\ref{A11}, $\tilde{\mathcal{A}}_{11}=0$. Here, $\tilde{\mathcal{A}}_{11}$ is the appropriate matrix element of the full transfer matrix, $\tilde{\mathcal{A}}$ and the condition dictates that the amplitude of the associated wavefunction vanish outside the length of the wire.

\begin{figure}[]
\subfloat[][]{\includegraphics[width=0.5\textwidth]{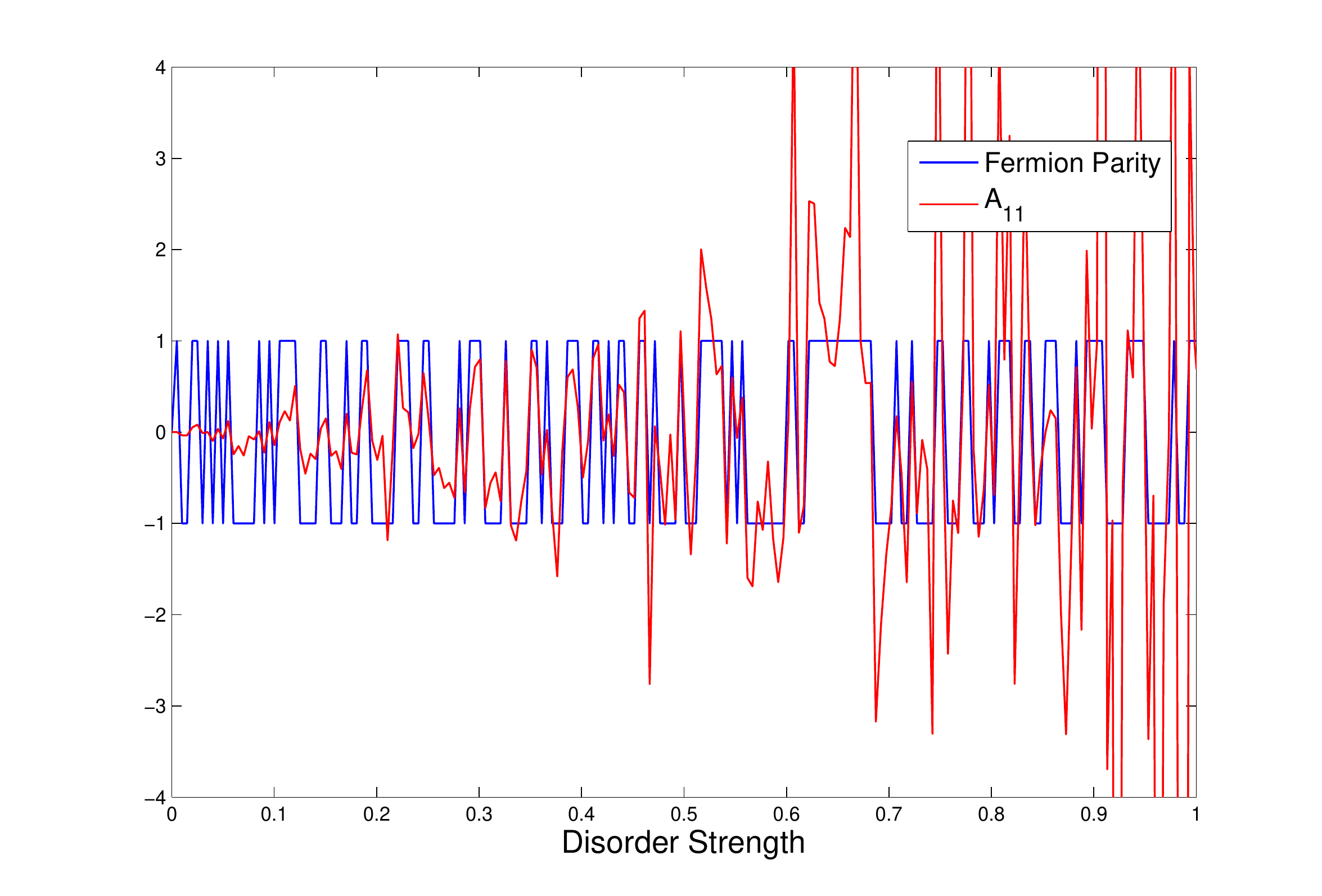}}

\caption{ Comparison between the fermion parity in a disordered Kitaev chain and the transfer matrix element $\tilde{\mathcal{A}}_{11}$ as a function of disorder strength(box disorder). The vanishing of the matrix element reflects the existence of a zero-energy Majorana state and can be seen here to coincide with parity switches as with the uniform case of Fig.\ref{fig:betaparity}.  Here $N=40,\Delta=0.6$}
 \label{fig:TMparity}
\end{figure}

In Fig.\ref{fig:TMparity}, we plot the matrix element $\tilde{A}_{11}$ as well as the ground state parity for a finite sized disordered wire as a function of disorder strength. Here, we use the Pfaffian measure of Eq. \ref{detB} for determining the parity. The points of vanishing $\tilde{\mathcal{A}}_{11}$ correspond to points which host a generic zero-energy state. Note that this state also has a zero energy partner, whose transfer matrix can be derived from the initial transfer matrix by replacing $\Delta$ with $-\Delta$. While the magnitude of $\tilde{\mathcal{A}}_{11}$ is unimportant for parity switch physics, its increase with disorder strength reflects the increase of the increase of the average energy splitting. Most prominently, we see that the vanishing of $\tilde{\mathcal{A}}_{11}$ is concurrent with the switching of ground state parity. Thus zero energy crossings correspond to the presence of two decoupled Majorana modes and associated degenerate parity states.

{\it In summary, the underlying normal Anderson model dictates zero-energy crossings in the disordered superconducting wire. These zero-energy crossings are exclusively associated with Majorana mode physics and correspond to the points when the system encounters a parity degeneracy in the process of undergoing a ground state parity switch.}


{\it Tracking parity switches in the topological phase diagram. ---}
 In obtaining a map of the regions where parity switches occur in the disordered Kitaev chain, we find that our studies from previous sections of the parity distributions in the uniform system serves as a guide. The precise points in parameter space where the switch occurs depend on the particular realization of disorder and are thus random. However, the parity switch phase diagram for the pure case in Fig. \ref{fig:paritydiagram} identifies the broad regimes in which parity switches can or cannot occur. In essence, for a fixed value of the the superconducting gap,  anywhere in the uniform chain phase diagram, windows in chemical potential where no parity switch occurs determine the width that the disorder distribution can span in the disordered case before a parity switch occurs. As we explicitly demonstrate, those observations allows us to chart out regimes where parity switches occur or not in the disordered wire. 

\begin{figure}[]
\centering 
\subfloat[][]{\includegraphics[width=0.3\textwidth]{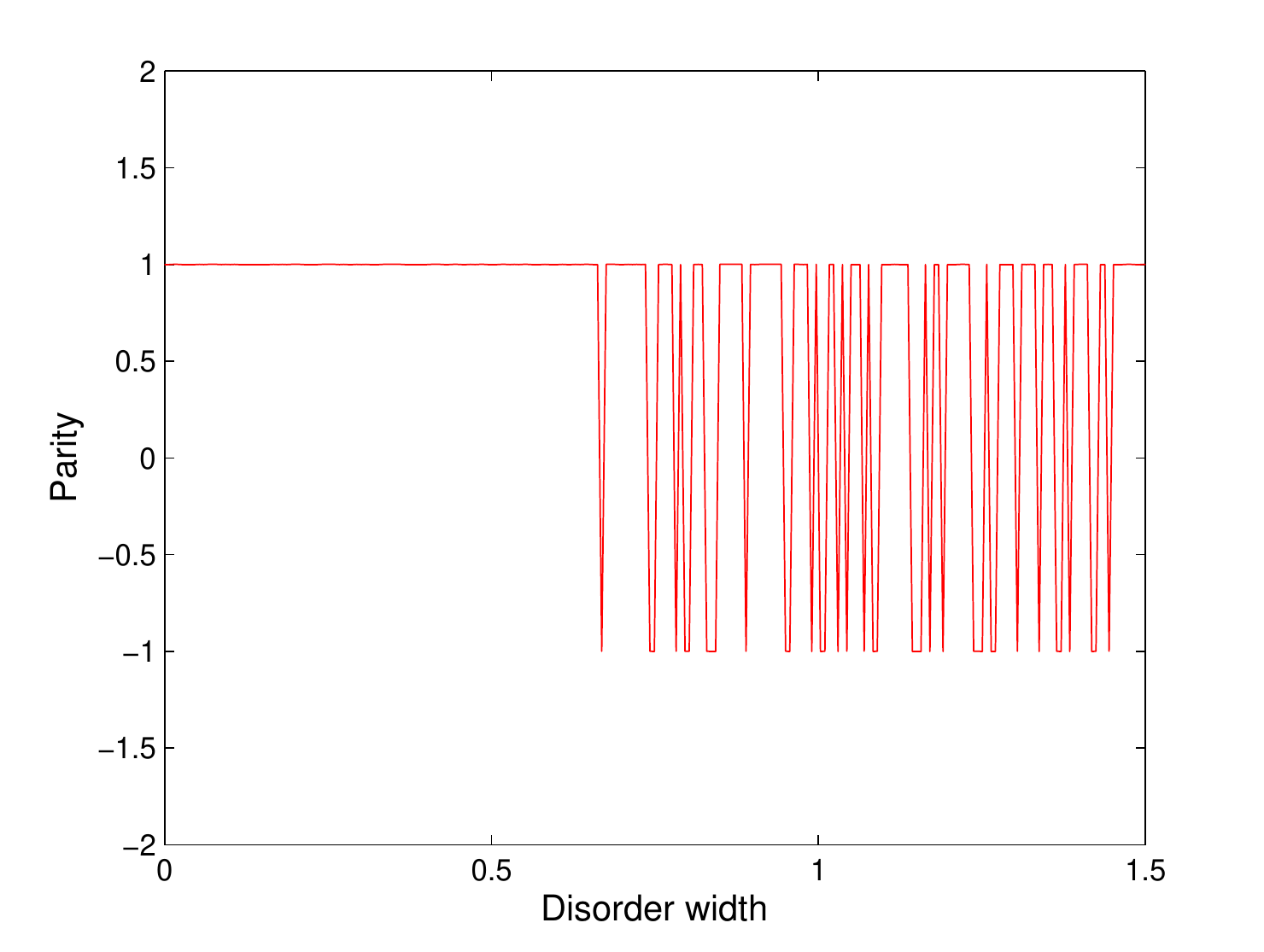}}\\
\subfloat[][]{\includegraphics[width=0.3\textwidth]{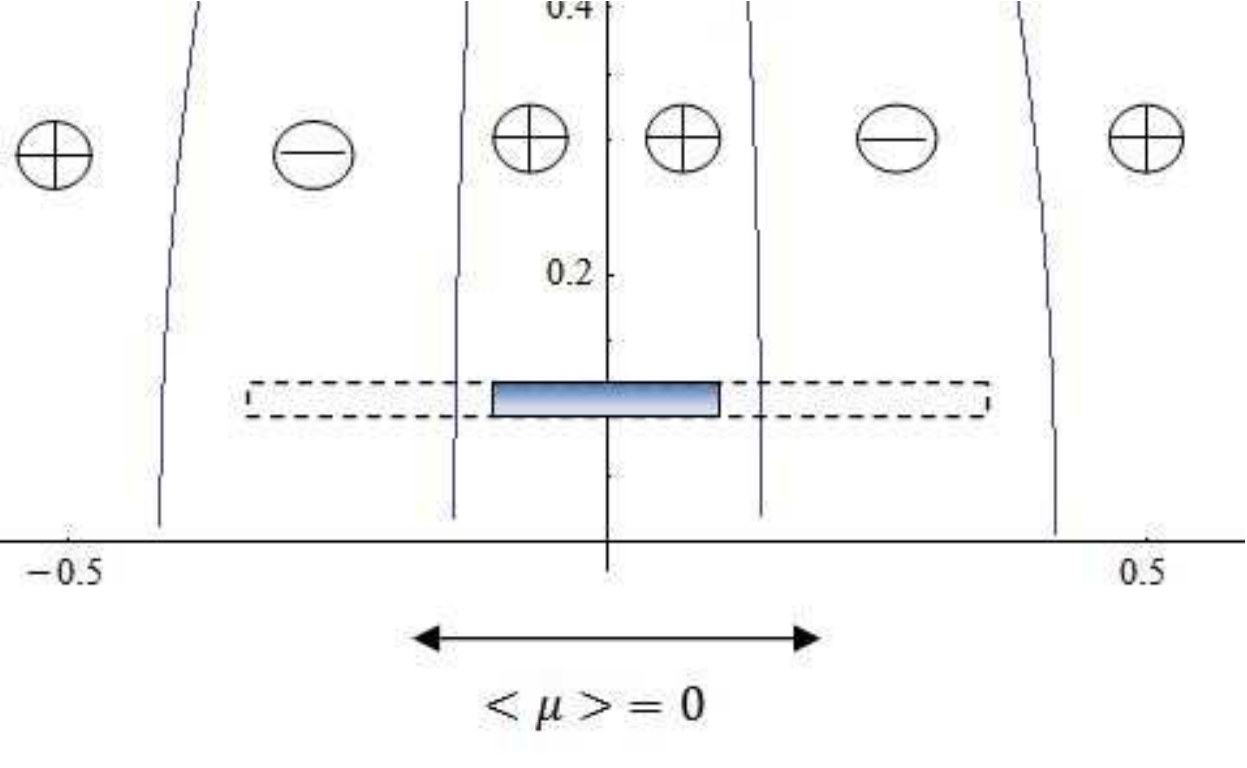}}\\

\caption {(a) Parity switches in a wire of an even number of lattice sites, $N=10$, as a function of disorder width for box disorder. Here, $\Delta=0.6$. Since parity is an even function of chemical potential for even $N$, the initial disorder window lies within a fixed parity as indicated in the uniform chain phase diagram in (b). As the disorder window is increased beyond a length-dependent value $\mu^p_{switch}$ of Eq. \ref{mucross} (dotted line) to include opposite parity sectors, parity switches begin to occur.}
 \label{fig:evendisorderswitch}
\end{figure}

  We first analyze parity switch behavior in chains having an {\it even} number of lattice sites.
 Figure ~\ref{fig:evendisorderswitch}.a shows numerical results for typical parity switching behavior as a function of disorder strength $W$. In all numerical simulations,  the values of $\mu_n$ are chosen randomly from a box distribution or a `window', centered at a mean $<\mu_n>$ value and having a width $W$. The parity is calculated again using the Pfaffian expression given by Eq. \ref{detB}.

One can see from Fig.\ref{fig:evendisorderswitch}.(a) that for the case of an even number of sites no parity switches occur up to a specific disorder window width. Beyond this width, switches start occurring in rapid succession and follow a random pattern that depends on the specific realization of disorder. As the disorder window width increases and includes opposite parity sectors, number of parity switches increase from being sparse to very dense.
 
 A qualitative picture for the parity switches can be obtained by invoking the properties of the uniform chain phase diagram. As shown in Fig.\ref{fig:evendisorderswitch}.(b) and discussed in previous sections, for a fixed wire length, the uniform chain ground state parity changes in a characteristic manner as a function of chemical potential. The chemical potential values at which these parity crossing happen are given by 
\beq
\mu_{switch}^p= 2\sqrt{1-\Delta^2} \cos\bigg(\frac{\pi p}{N+1}\bigg),
\label{mucross}
\eeq 
where $p$ takes integer values from $1$ to $N/2$ for even $N$. The first crossing occurs for $p=N/2$ and the width of the central parity sector is $2 \mu_{switch}^{N/2}$. Thus, any value of chemical potential lying within this window is associated with the same parity. Upon introducing disorder such that the site-dependent chemical potential lies within this width, we would expect no changes in the overall ground state parity. Beyond this width, however, chemical potentials associated with the opposite parity become included in the on-site distribution, allowing for the possibility of a global ground state parity switch. The probability of such a switch increases with increasing disorder window width as it allows a higher chance of on-site chemical potentials being associated with opposite parity. This qualitative picture is consistent with the behavior of parity switches in Fig. \ref{fig:evendisorderswitch}.b. We now show that it accounts for our numerical findings with regards to parity switch behavior as a function of system size, average chemical potential off-set, and odd versus even number of lattice sites.

 The occurrence of parity switches highly depends on the system size, $N$. As can be seen in Eq. \ref{mucross}, as $N$ increases, the chemical potential value at which the first parity switch occurs in the uniform case, $\mu_{switch}^{N/2}$, becomes smaller. Thus, in the disordered case, for a wire of longer length, we expect parity switches to commence for smaller disorder window width. We indeed find this to be true.

\begin{figure}[]
\centering 
\subfloat[][]{\includegraphics[width=0.3\textwidth]{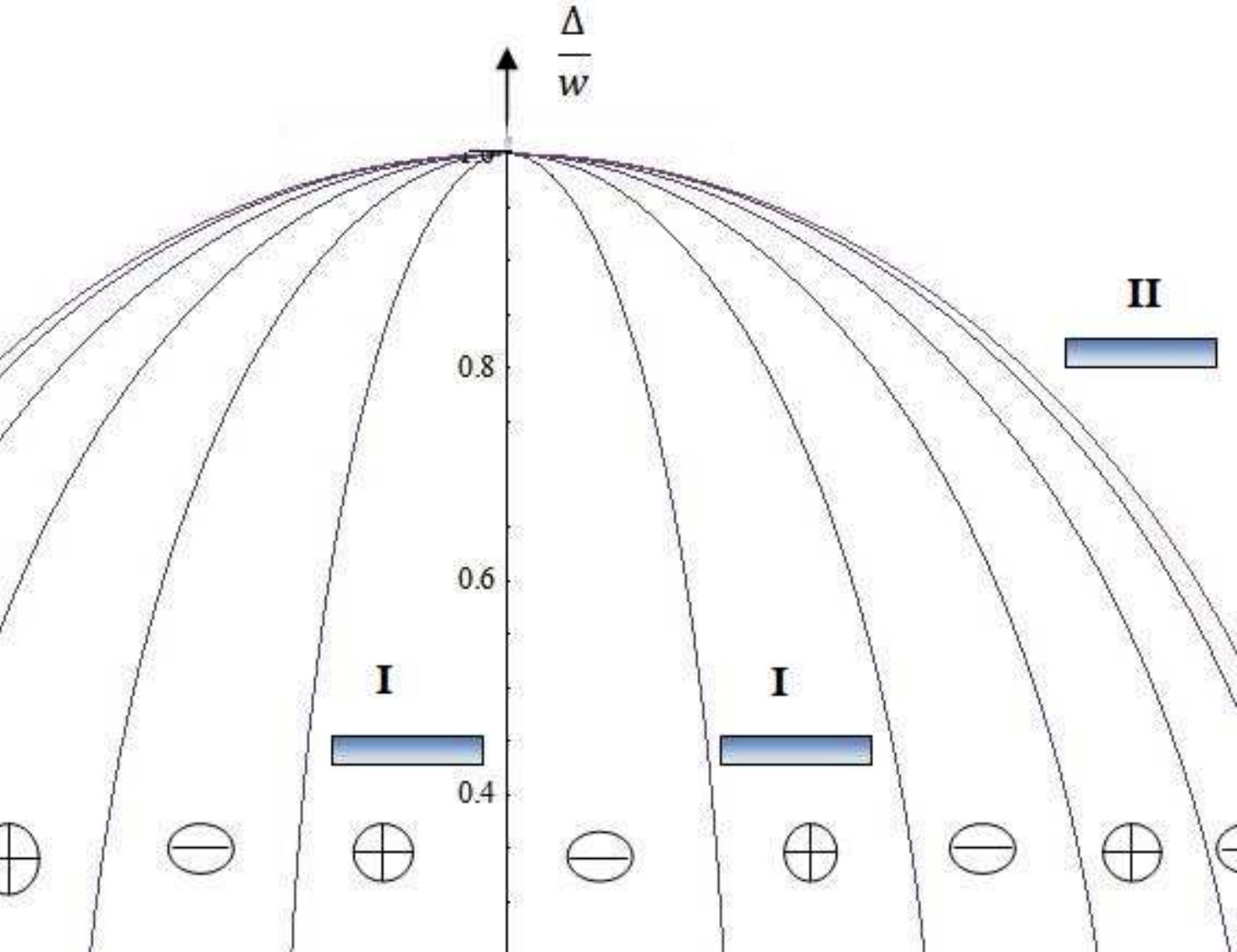}}\\
\subfloat[][]{\includegraphics[width=0.3\textwidth]{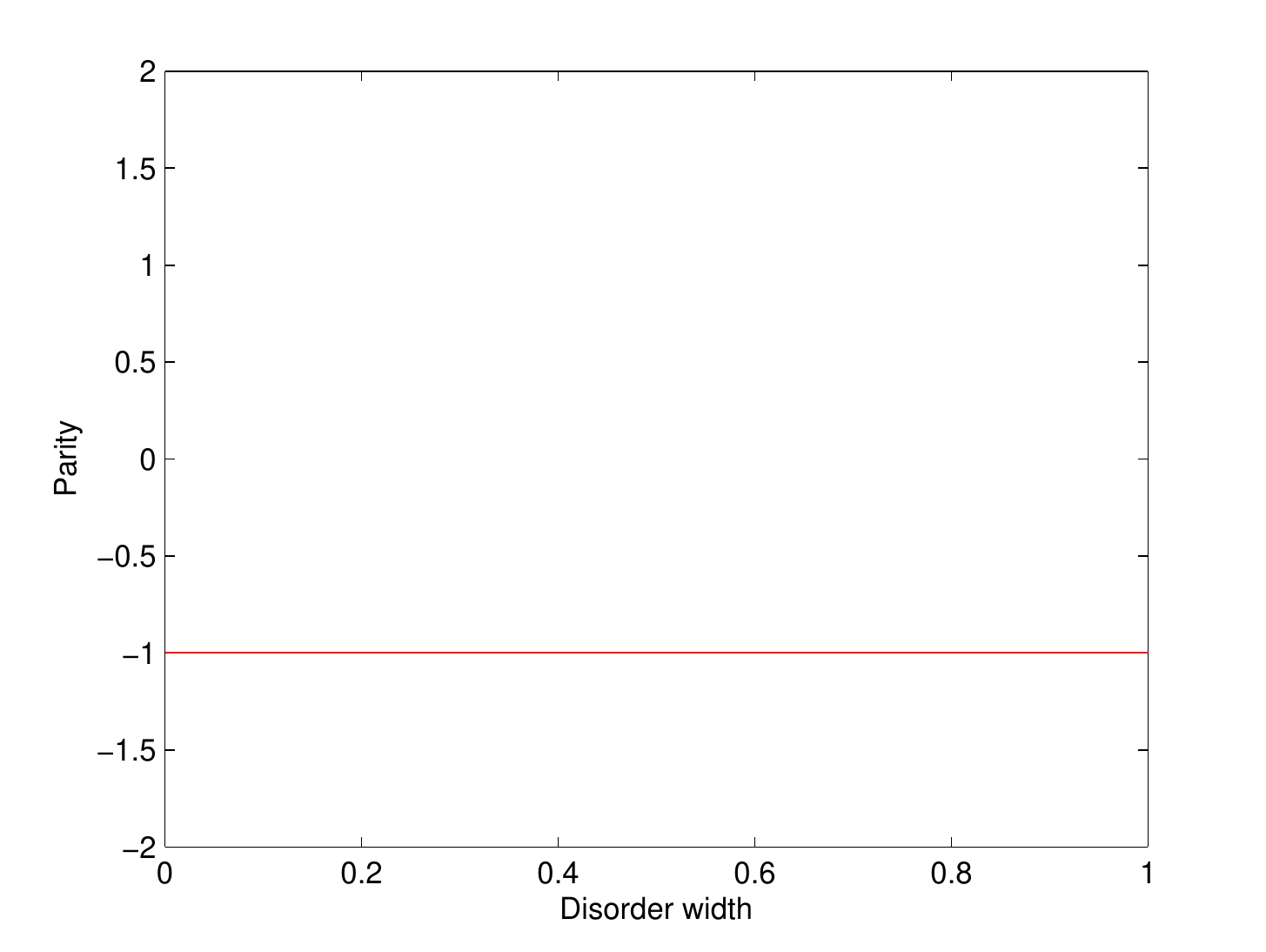}}

\caption {Possibility of no parity switches in the disordered Kitaev chain.(a) Two possible disorder distributions  centered around finite chemical potential are shown in which parity switches are not expected - I. Double box disorder in two disjoint sectors of the same parity and II. Box disorder outside the circle of oscillations. (b) In both cases, the plot of parity as a function of disorder strength indeed shows the absence of parity switches. }
 \label{fig:disordernoswitch}
\end{figure}

{\it Chemical potential off-set and absence of parity switches. ---} 
As with the case above of disorder centered around zero chemical potential, here we analyze parity switches in the presence of a chemical potential off-set where the mean $<\mu_n> \neq 0 $. Once again, the uniform chain parity regimes inform the behavior of the disordered wire. One of the most striking features to emerge is that if the chemical potential off-set and disorder window width are chosen to lie within a region of the uniform chain phase diagram where no parity switches occur, then the disordered wire too shows no ground state parity switch. Specifically, from Eq. \ref{mucross}, we see that the chemical potential span of any given parity sector in the uniform wire is given by 
\bea \non
\Delta \mu_p = 
2\sin\bigg(\frac{\pi (p+1/2)}{N+1}\bigg) \sin\bigg(\frac{\pi}{N+1}\bigg)
\eea 
This sector width shrinks with increasing $N$. Also for a given size $N$, this width decreases as one progresses from zero chemical potential to the boundary of the circle of oscillation (COO), i.e. as $p \rightarrow 1$. Furthermore, outside the COO, no parity switches occur. These features provide bounds for the values of chemical potential off-set and disorder window width in which no parity crossings occur.

In Fig.\ref{fig:disordernoswitch}, we explicitly verify the observations made above. As one example, we consider a double box disorder distribution where two disorder windows are centered around two values of chemical potential such that both windows lie within the same parity sectors in the uniform chain phase diagram. As a second example, we place the disorder window outside the COO, thus expecting no parity switches. Indeed, in both situations parity switches are not observed.  While the double box disorder seems unrealistic, the case of taking the window outside the circle might be possible to realize experimentally. One can study the `Ising-limit' $\Delta=1$, which is tangential to the circle. In this case taking $<\mu_n>=0$, the parity switches as a function of width again depends on the number of sites being even or odd. This explicitly shows that there are cases where one need not have any parity switches even in the presence of disorder and when they are present, the behavior completely depends crucially on the features of the parity sectors of the uniform Kitaev chain.

\begin{figure}[]
\centering 
\subfloat[][]{\includegraphics[width=0.3\textwidth]{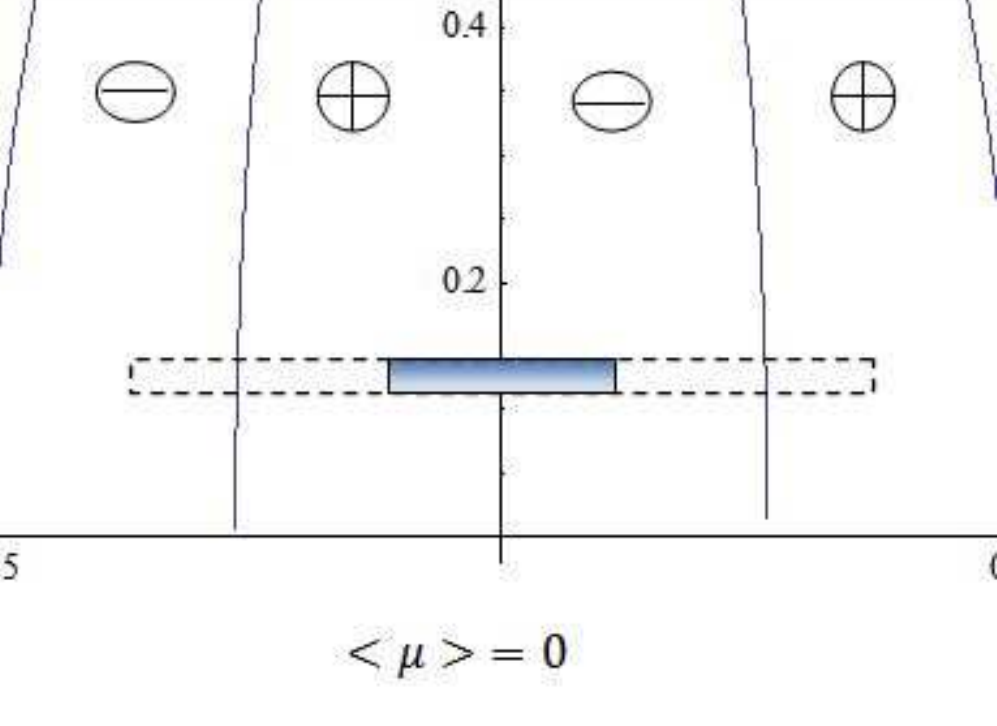}} \\
\subfloat[][]{\includegraphics[width=0.3\textwidth]{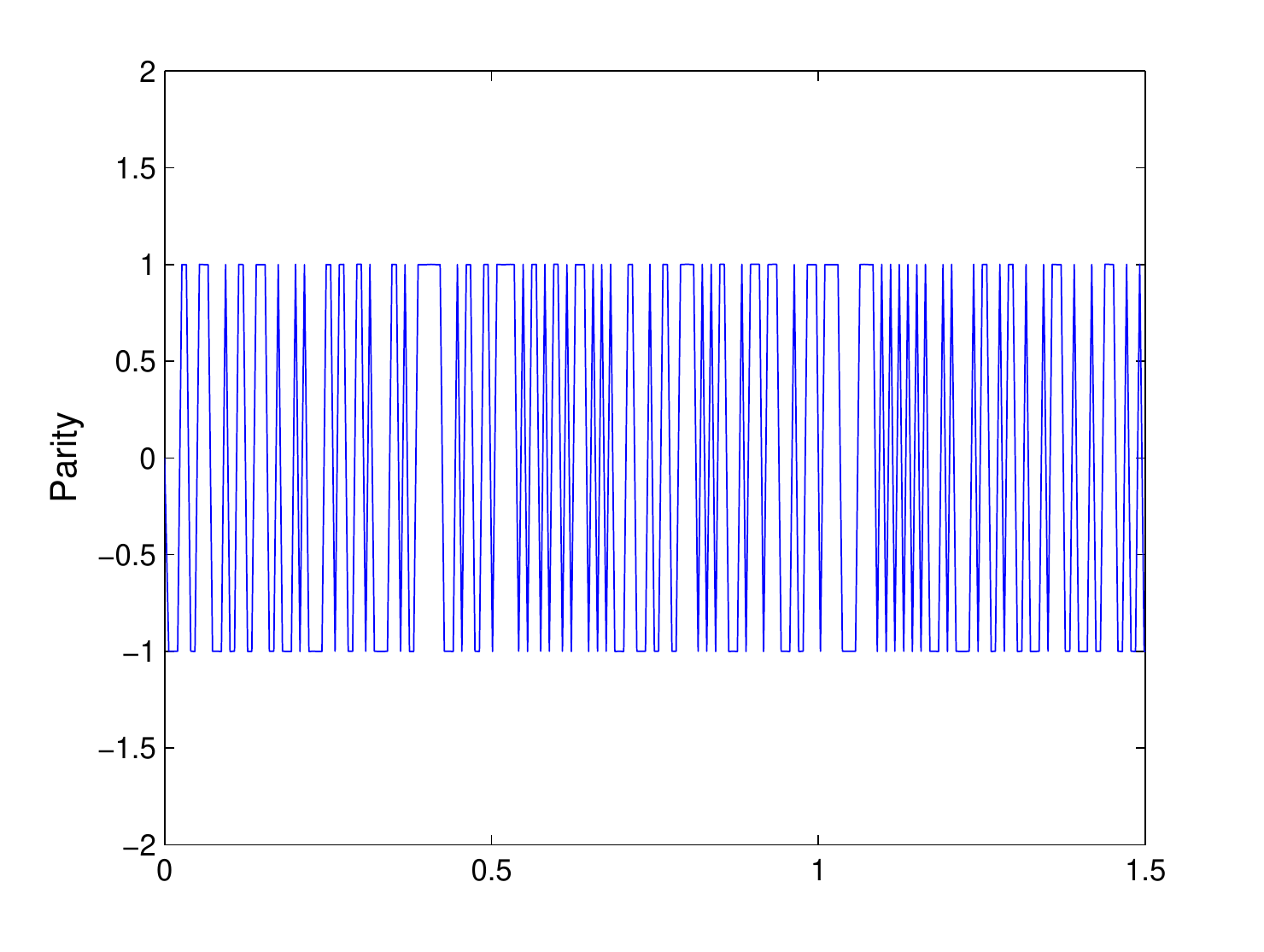}}

\caption { Parity switches in a wire of odd length ($N=11, \Delta=0.6$) as a function of disorder strength for box disorder. (a) Given the  antisymmetry in parity sectors across $\mu=0$ for the uniform case, the slightest change in disorder strength is expected to produce a parity switch. This is confirmed in (b), which shows a profusion of random parity switches as a function of disorder strengths starting from the smallest amount of disorder.}
 \label{fig:oddparitydisorder}
\end{figure}

{\it Dependence on even versus odd $N$. --- } A noteworthy difference arises in the behavior of parity switches between chains of even and odd number of lattices sites, once again stemming from the structure of the uniform wire ground state parity distribution. The key difference is that the distribution of parity switches is symmetric as a function of chemical potential for the even site case but anti-symmetric for the odd site case. Explicitly, the values of chemical potential for when parity crossing occur are given by Eq. \ref{mucross} for an even number of sites. While the same holds for the odd site case (with values of $p$ ranging from 1 to $(N-1)/2$), a parity switch also occurs at $\mu=0$. Thus, in the presence of disorder,  in contrast to the even site case shown in Fig. \ref{fig:evendisorderswitch}, even the narrowest disorder window centered around zero chemical potential gives rise to parity switches. This behavior is corroborated in Fig. \ref{fig:oddparitydisorder}.
 
 Thus, we have presented a qualitative study of ground state parity in finite sized disordered Kitaev chains. Salient features are that the uniform Kitaev chain serves to inform parity switches in disordered chains. The transfer matrix, as with the uniform case, tracks zero energy crossings and shows that they are consistent with parity switches, thus attributing all such crossings with Majorana mode physics. For the disordered case, the underlying normal state Anderson problem determines oscillations of the Majorana wavefunctions and associated parity switches. Finite length analyses of the uniform wire have direct bearing on parity switching behavior for the disordered case. Windows in chemical potential that contain fixed parity sectors in the uniform case provide bounds for disorder distribution widths that respect no parity switching. Specifically, this observation results in the characteristic parity switching behavior for even and odd length chains shown in Fig.\ref{fig:evendisorderswitch} and Fig.\ref{fig:oddparitydisorder} as well as regimes in the phase diagram where no parity switching takes place, as shown in Fig.\ref{fig:disordernoswitch}.

\section{Discussion}
\label{Conclude}
To summarize our approach and results, we have presented a detailed analysis of the Majorana wavefunctions and fermion parity switches in the finite-sized Kitaev chain. Employing the Majorana transfer matrix and studying its properties enabled us to resolve the effect of any general potential landscape on the Majorana wavefunction. The wavefunctions are characterized by decay and oscillations where we found that the latter stems purely from the underlying normal tight-binding model. In the uniform system, these oscillations correspond to band oscillations which, when analyzed as a function of parameter space, allowed us to identify a circular regime in the topological phase in which Majorana wavefunctions oscillate. In the disordered case, the underlying normal tight-binding model is the Anderson model. Thus the underlying oscillations are random oscillations stemming from the Anderson problem and there are large fluctuations between different disorder realizations.  Any further investigation on the oscillations and the resulting parity flips can thus bank on the vast literature on disorder and Anderson localization. 

  The disordered one-dimensional p-wave superconducting wire has been studied extensively in the past for its rich localization and Majorana physics. On this front, our work performs one of the initial studies probing fermion parity effects in this system. We show that in spite of the random nature of the parity switches in the presence of disorder, the parity sectors of the uniform Kitaev chain can still dictate various qualitative features of switches in the disordered case.  One of the striking observations in our studies is the presence of regimes in which no parity switching takes place for a range of disorder strengths. These features can have strong bearing on realistic protocols for topological quantum computation, where the operations are through ground state parity manipulations. 

 Our study here opens up several future directions concerning Majorana wavefunction features and parity switches. With regards to disorder, several aspects can potentially be studied by invoking known results from the literature on disordered spin chains. One challenge with characterizing topological features in the presence of disorder, just as with non-topological features, is that certain quantities of interest might show large sample-to-sample fluctuations and appropriate quantities need to be identified for disorder averaging. For example, in the case of Anderson localization, the conductance itself has large fluctuations whereas the logarithmic conductance shows a Gaussian distribution in certain limits. Specifically, seeking a quantity for cleanly characterizing parity switches is in order; related studies in the context of Josephson junctions have been performed using random matrix theory ~\citep{Beenakker13, Beenakker13A}.

  Our treatment here enables the future study of a range of potential landscapes. A crucial feature in our work here and in previous ones is that we map features of the Majorana bound mode through the behavior of the underlying normal tight-binding problem. In light of this observations, a specific case of interest would be that of  (quasi-)periodic potentials. The wave equation for Majorana modes in this case respects Harper's equation, which has been studied extensively for its mathematical richness. As already pointed out in Ref. \onlinecite{DeGottardi13}, the Lyapunov exponent reflects the Hofstadter  butterfly pattern and we expect Majorana wavefunctions and parity switches to do the same.

  On the front of extensively studied realistic systems, our treatment directly applies to proposed and experimentally investigated systems  such as  proximity induced spin-orbit-coupled wires and superconductor-topological insulator hetero-structures. In these systems, the Kitaev chain forming the basis of our work provides an excellent prototype and we expect our observations on  Majorana oscillations and parity switches to be directly applicable. A possible avenue for exploring parity effects would involve coupling the Majorana wire  to a charge sensitive system, such as a quantum dot, a single electron transistor, or a scanning tunneling microscope(STM) tip ~\cite{Flensberg12, higginbotham15, zocher13, liu14, da14, dai15, Ben-Shach15}. Other methods which have gained prominence in the Majorana mode context include coupling to microwave cavity, circuit quantum electrodynamics, cooper pair boxes and Transmon systems seem to show promise and have gained much attention, especially as they directly couple to the parity sectors arising from the Majorana modes~\citep{Yavilberg15, Ginossar14, Dmytruk15, deLange15, Ohm15, Vayrynen15,Hell16, Knapp16}. There have also been several proposals for realization of topological systems hosting Majorana modes in cold atomic systems ~\citep{Mazza12, Jiang11, Buhler14}. These systems and probes put together offer ample ways of accessing and exploring the wavefunction oscillations and parity regimes delineated here. 
  
  Having a precise handle of the parity landscape in Majorana wires is one of the key requirements for several topological quantum computational considerations. As an example, as shown in Refs.\onlinecite{Bonderson08, Bonderson13, Burello13, Sau11, Burnell14}, one method of performing non-Abelian rotations in the degenerate Majorana ground state manifold is through tuning the coupling between Majorana modes. This method has formed the basis of various quantum computational protocols~\cite{Alicea11, Sau11, vanHeck12, Hassler11, Bonderson08, Bonderson13, Knapp16}. Thus the detailed knowledge presented here on the degeneracy points in the topological phase diagram, parity switches in a finite length wire, and their behavior in the presence of disorder are all of relevance in this context.

Finally, an obvious extension of our work would be to include interactions in the system. It has been shown using Luttinger liquid treatments interaction of small interaction strength would not destroy the topological phase as long as the superconducting gap is maintained in the system  ~\citep{Lutchyn11, Sela11, Gangadharaiah11, Stoudenmire11, Goldstein12, Yang14, Chan15, Lobos12, Deb16}. The issue now would be to cast parity effects in terms of a many-body generalization of the non-interacting case ~\citep{Kells15, Kells15A}. A possible approach would entail drawing from the exact map between the Kitaev chain Hamiltonian with nearest neighbor density-density interaction and the transverse-field XYZ Heisenberg spin chain~\citep{Gergs16, Katsura15, Fendley15}. Such systems with both disorder and interactions have also been studied actively in the recent years in the context of many-body localization. These considerations regarding interactions, along with several other avenues, will form the basis of further studies on wavefunction oscillations and fermion parity switches in Majorana wires.

\begin{acknowledgements}
We are grateful to Diptiman Sen, Wade DeGottardi and Victor Chua for illuminating 
conversations. This work is supported by the National Science Foundation 
under the grants DMR 0644022-CAR and the U.S. Department of Energy, Division of Materials Sciences under Award No. DE-FG02-07ER46453.
\end{acknowledgements}

\bibliographystyle{apsrev}
\bibliography{DisorderPaperv3}
\end{document}